\documentclass[sigconf]{acmart}

\usepackage{mathtools,amsthm}
\usepackage{nicefrac}
\usepackage{subcaption}
\usepackage{tikz}
\usepackage{xfrac}
\usetikzlibrary{quantikz}

\def\hush#1{}

\AtBeginDocument{%
  \providecommand\BibTeX{{%
    \normalfont B\kern-0.5em{\scshape i\kern-0.25em b}\kern-0.8em\TeX}}}

\setcopyright{none} 
\renewcommand\footnotetextcopyrightpermission[1]{} %removes footnote with conference information in first column
\hush{
\setcopyright{acmcopyright}
\copyrightyear{2020}
\acmYear{2020}
\acmDOI{}
}

\acmConference[]{}{}{}

\usepackage{color, colortbl}
\usepackage{hyperref}
\usepackage{graphicx}
\usepackage{listings}

\definecolor{dkgreen}{rgb}{0,0.6,0}
\definecolor{gray}{rgb}{0.83,0.83,0.83}
\definecolor{mauve}{rgb}{0.58,0,0.82}

\newcommand{\Xsqr}{$\mathit{X}^{\sfrac{1}{2}}$}
\newcommand{\Ysqr}{$\mathit{Y}^{\sfrac{1}{2}}$}

\newcommand{\gateH}{\gate[style={fill=blue!30}]{H}}
\newcommand{\gateT}{\gate[style={fill=gray}]{T}}
\newcommand{\gateZ}{\gate[style={fill=yellow!50}]{Z}}
\newcommand{\gateX}{\gate[style={fill=teal!30}]{X}}
\newcommand{\gateY}{\gate[style={fill=purple!30}]{Y}}

\def\blue#1{\textcolor{blue}{#1}}

\begin{document}

\title{Faster Schr\"odinger-style simulation of quantum circuits}

\author{Aneeqa Fatima}
%\authornote{}
\email{aneeqaf@umich.edu}
\affiliation{%
  \institution{University of Michigan}
  \streetaddress{2260 Hayward St.}
  \city{Ann Arbor}
  \state{MI}
  \postcode{48109-2121}
}

\author{Igor L. Markov}
\email{imarkov@umich.edu}
\affiliation{%
  \institution{University of Michigan}
  \streetaddress{2260 Hayward St.}
  \city{Ann Arbor}
  \state{MI}
  \postcode{48109-2121}
}
%\renewcommand{\shortauthors}{Fatima and Markov}

\begin{abstract}
   Recent demonstrations of superconducting quantum computers by Google and IBM and
   trapped-ion computers from IonQ fueled new research in quantum algorithms, compilation into quantum circuits,
and empirical algorithmics. While online access to quantum hardware remains too limited to meet the demand, simulating quantum circuits on conventional computers satisfies many needs. We advance Schr\"odinger-style simulation of quantum circuits that is useful standalone and as a building block in layered simulation algorithms, both cases are illustrated in our results. Our algorithmic contributions show how to simulate multiple quantum gates at once, how to avoid floating-point multiplies, how to best use data-level and thread-level parallelism as well as CPU cache,
and how to leverage these optimizations by reordering circuit gates. While not described previously, these techniques
implemented by us supported published high-performance distributed simulations up to 64 qubits. To show additional impact, we benchmark our simulator against Microsoft, IBM and Google simulators on hard circuits from Google.
\end{abstract}


\hush{
}

\hush{
}



\maketitle

\section{Introduction}

Quantum computation was first developed theoretically to accelerate computational bottlenecks using quantum-mechanical phenomena \cite{NC}. Among several promising quantum models of computation, quantum circuits have been implemented in several technologies, for which end-to-end programmable computation has been demonstrated at intermediate scale \cite{Preskill}. In 2019, Google claimed quantum-computational supremacy~\cite{GoogleSuprem,Arute2019} by scaling quantum computation to the point where simulating it becomes exceptionally challenging even on supercomputers. The significance of quantum design automation tools has been appreciated
for many years, to the point where IBM and Microsoft have developed extensive toolchains
(QISKit and QDK respectively), which include language support, compilation, optimization,
and a variety of execution back-ends for physical quantum computers and circuit simulators.
Just like in conventional Electronic Design Automation, such software allows one to validate
prototype designs before building the hardware.

\noindent {\bf Quantum circuit simulation} in general is inherently difficult due
to the exponential growth in the number of internal parameters and complexity-theoretic reasons \cite{aaronson2016complexity,bouland_quantum_2018}. We distinguish several categories and uses of quantum-circuit simulation:
\begin{enumerate}
\item Polynomial-time simulation of "easy" special-case quantum circuits --- those using Clifford gates \cite{AG_2004}, many instances of
Grover's algorithm \cite{Viamontes}, and circuits with small tree-width \cite{TNC2008}. Such simulation is used ($i$) to rule out quantum speed-up in specific algorithms, and ($ii$) for limited initial testing of quantum computers in the lab and debugging of failed tests.
\item Best-effort simulation of unrestricted "small" quantum circuits up to 40 qubits, as in Section \ref{sec:empirical}, and also error modeling. Such simulation is used ($i$) for broad testing and routine debugging of quantum computers, 
($ii$) to verify local quantum circuit transformations and whole circuits, as well as
($iii$) to evaluate new quantum algorithms, quantum error-correcting codes and device architectures~\cite{QS_opportunities}.
In particular, VQE algorithms common for quantum-chemistry applications run numerous quantum circuits in a sequence, their development particularly benefits from fast simulation \cite{quantumChemistry_2020}.
\item Distributed quantum-circuit simulation on
clusters and supercomputers \cite{de_raedt_massively_2007,de_raedt_massively_2018,PB,qHIPSTER,IBM2017breaking,IBM2019storage,Villalonga_2020}, including the world's largest \cite{Taihu}. Such expensive and scarce resources are used to verify quantum computation and set performance baselines when claiming quantum-computational supremacy \cite{GoogleSuprem,Arute2019,Villalonga_2020}. Whereas other uses
entail repeated on-demand simulations on readily available computing hardware, this category targets a small number of "expensive" one-off simulations.
\end{enumerate}
All three categories of quantum-circuit simulation are in demand today, but high-performance simulation of unrestricted quantum circuits in Categories 2 and 3 is challenging and motivates algorithmic improvements of the type we propose.
Using fast simulation to evaluate, verify, test and debug
larger quantum circuits and algorithms facilitates the development of many applications.

\noindent
{\bf Schr\"odinger-style simulation} is
the mainstream technique for general-case simulation of quantum algorithms, circuits and physical devices. It represents a quantum state (wave function) by a vector of complex-valued amplitudes and modifies this vector in place by applying quantum transformations (quantum gates, laser pulses, algorithmic modules, etc). Schr\"odinger simulation
\begin{itemize}
    \item scales linearly with computation (circuit) depth but exponentially with the number of qubits, or width (Section 
    \ref{sec:scalability});
    \item is commonly used for small and mid-size quantum-circuit and device/technology simulations because its unoptimized variants are relatively straightforward to implement;
    \item dominates supercomputer-based quantum circuit simulations because it can leverage distributed memory, fast interconnect, GPUs, etc \cite{de_raedt_massively_2007,wecker2014liquid,de_raedt_massively_2018,PB,qHIPSTER,IBM2017breaking,Taihu,IBM2019storage};
    \item has been extended for better scalability via {\em layered simulation} \cite{rr2020,teleport2020,IBM2019storage}. For example, combining Schr\"odinger simulation with Feynman path summation enables scaling tradeoffs between circuit depth and width \cite{aaronson2016complexity}
    and orders-of-magnitude resource savings in important cases
    \cite{rr2020}. 
\hush{
    .\footnote{A recent implementation of such Schr\"odinger-Feynman simulation on the Google Cloud \cite{rr2020} reports orders-of-magnitude reductions in resource requirements compared to prior supercomputer simulations in \cite{PB}.}
}
\end{itemize}
 Our work accelerates both pure Schr\"odinger simulation and layered algorithms that use it, as we illustrate empirically for Schr\"odinger-Feynman simulation from \cite{rr2020}. 
Our algorithmic insights and innovations offer both constant-time implementation speed-ups and algorithmic speed-ups that scale with qubit count:

\begin{itemize}
\item Numerical accuracy improvements and checks to enable a compact floating-point data type which reduces memory footprint and bandwidth,
and facilitates SIMD instructions.
\item Avoiding most floating-point multiplications in favor of faster additive and bitwise instructions.
\item Batched simulation of diagonal quantum gates that significantly reduces expensive memory traversals and the overall runtime while exposing thread-level parallelism.
\item Encoding sets of same-type diagonal gates by bitmasks (Figure \ref{fig:CZ}), simulating them with bitwise and mod-$p$ CPU instructions, and speeding up simulation with Gray codes.
\item Encoding sets of same-type single-qubit (not necessarily diagonal) gates by bitmasks (Figure \ref{fig:reorder}) and simulating them using a recursive FFT-like algorithm that improves cache locality and exposes thread-level parallelism.
\item The insight that some quantum gates (implemented in superconducting quantum computers)
are easier to simulate in pairs because this simplifies matrix elements and benefits from
batched load/store operations.
\item {Gate clustering \em by type}, contrasted with the common {\em gate fusion} that clusters heterogeneous gates that share qubits. We develop a reordering-based clustering algorithm that finds larger homogeneous clusters (Figure \ref{fig:reorder}).
\item Aligned memory read-write operations and {\em gate clustering by cache line}. Here we read entire L1 cache lines from memory and apply multiple gates to a cache line when possible.
\item Implementing our algorithms 
with extensive use of AVX-2 instructions that
improves productivity per instruction.
\end{itemize}

Our empirical results start with comparisons on smaller circuits where software from IBM and Microsoft can be used, then study the scalability of our techniques and show how they boost layered simulation methods. 
 {\em On a MacBook Pro laptop} with 16GiB RAM, we simulate circuits with a $5 \times 5$-qubit array to any depth, with $20\times$ and $12\times$ speedups over simulators from Microsoft QDK and IBM QISKit / QASM, respectively. Our simulator Rollright uses $3.27\times$ less memory than QDK and can simulate $6\times5$-qubit circuits of any depth. 
 {\em On a midrange server}, we simulate up to $6\times 6$-qubit circuits and illustrate Schr\"odinger-Feynman simulation \cite{rr2020} 
 that combines half-sized Schr\"odinger simulations, benefits from our techniques, and shows up to $4000\times$ speedups over QDK and earlier versions of QISKit. Profiling data, ablation experiments, as well as comparisons to Google Qsim showcase the impact of specific innovations.

 In the remaining part of the paper, Section \ref{sec:background} gives minimal background in quantum circuits and circuit simulation. Section \ref{sec:baseline}
 outlines baseline Schr\"odinger-style simulation. Our algorithmic framework is presented
in Section \ref{sec:framework}, including design decisions and some performance optimizations. In Section \ref{sec:HW}, we leverage the CPU architecture and hardware resources. Empirical results and scalability are reported in Section \ref{sec:empirical}, with conclusions in Section \ref{sec:conclusions}.
%Key concepts are illustrated by examples.

%\vspace{-5mm}
\section{Background}
\label{sec:background}

 A {\em quantum circuit} on $n$ qubits is a sequence of {\em quantum gates} that act on {\em quantum states} represented by $2^n$-dimensional complex-valued vectors \cite{NC}.
The computation usually starts with the basis vector $(1,0,\dots,0)$, sets each qubit 
 in the $\ket{0}$ state rather than the $\ket{1}$ state. Quantum gates transform this
 state into some superposition of $2^n$ basis vectors (each labeled by some $n$-bit binary number $j$ and participates with the complex {\em amplitude} $\alpha_j$).
 Quantum measurements are traditionally performed at the end to stochastically read out non-quantum bits, while destroying the quantum state.
 The probabilities of outcomes depend on $\alpha_j$. In this work, we assume sufficient memory to represent all $\alpha_j$ and seek to find them all ({\em strong simulation}). Simulating measurements is then straightforward.
 
 Industry computers use a handful of one- and two-qubit gate types,
 defined by $2\times2$ or $4\times4$ unitary matrices~\cite{NC}.
 Qubits are usually arranged in a planar grid, and two-qubit gates are restricted to nearest-neighbor qubits. Yet, our methods handle two-qubit gates acting on any pair of qubits (as in ion-trap computers).
 
 Single-qubit quantum gates include
\[
\mathit{NOT}= \begin{bmatrix} 0 & 1 \\ 1 & 0 \end{bmatrix},
H = \frac{1}{\sqrt{2}}\begin{bmatrix*}[r] 1 & 1 \\ 1 & -1 \end{bmatrix*} \text{, and } 
Z = \begin{bmatrix*}[r] 1 & 0 \\ 0 & -1 \end{bmatrix*},
\]
where $\mathit{NOT}$ negates the state of the qubit, $H$ sets the qubit into a superposition of $\ket{0}$ and $\ket{1}$, while $Z$ shifts the phase of the qubit.
Multiple qubits can be coupled using the Controlled-NOT ($\mathit{CNOT}$ and Controlled-$Z$ ($\mathit{CZ}$) gates:
\[
\mathit{CNOT} = \begin{bmatrix} 1 & 0 & 0 & 0 \\ 0 & 1 & 0 & 0 \\ 0 & 0 & 0 & 1 \\ 0 & 0 & 1 & 0 \end{bmatrix} ~~~~
\mathit{CZ} = \begin{bmatrix} 1 & 0 & 0 & 0 \\ 0 & 1 & 0 & 0 \\ 0 & 0 & 1 & 0 \\ 0 & 0 & 0 & -1 \end{bmatrix}.
\]

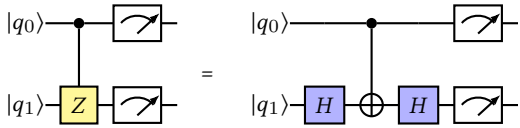
\begin{figure}
	\centering
	\begin{quantikz}[column sep=0.2cm]
        \ket{q_0}&  \ctrl{1} &  \meter{} & \qw \\
        \ket{q_1} & \gateZ &  \meter{} & \qw \\
    \end{quantikz}\hspace{0.3cm}=\hspace{0.3cm}\begin{quantikz}[column sep=0.2cm]
        \ket{q_0}& \qw & \ctrl{1} & \qw &  \meter{} & \qw \\
        \ket{q_1} & \gateH & \targ{} &  \gateH & \meter{} & \qw \\
    \end{quantikz}
    \vspace{-6mm}
	\caption{Quantum circuit diagrams for equivalent circuits}
	\label{fig:ckt}
	\vspace{-2mm}
\end{figure}

\begin{example}
 Figure \ref{fig:ckt} illustrates a two-qubit  circuit with two Hadamard gates around a $\mathit{CNOT}$ gate, followed by measurements on each qubit.
 The three-gate circuit is equivalent to a $\mathit{CZ}$ gate:
 \begin{equation}
  \label{eq:CZ}
(I\otimes H) ~ \mathit{CNOT} ~ (I\otimes H) = \mathit{CZ},
\end{equation}
 where $\otimes$ represents the Kronecker product
 and $I$ represents the identity matrix of appropriate dimension.\footnote{A similar equation expresses $\mathit{CNOT}$ via $\mathit{CZ}$ and two $H$ gates.} Given that the $\mathit{CZ}$ gate is diagonal, it maps $\ket{11}$ into  $-\ket{11}$ and the remaining three basis vectors to themselves. Therefore, if the circuit starts with any basis
 vector, the measurement will deterministically
 produce this vector. In general, diagonal operators/gates do not create superpositions.  In many circuits, one-qubit gates create {\em separable superpositions}, which diagonal gates then turn into {\em entangled superpositions}.
 \end{example}

Quantum circuits with gates shown above (along with $P=Z^{\sfrac{1}{2}}$) can be simulated in polynomial time by a compact algorithm, hence offer no quantum computational advantage \cite{AG_2004}. Google Bristlecone and Sycamore chips \cite{google_2018,GoogleSuprem,Arute2019} also support the following gates
\[
X^{\sfrac{1}{2}} 
= \frac{1}{2}\begin{bmatrix*}[r] 1+i & 1-i \\ 1-i & 1+i \end{bmatrix*}, ~
Y^{\sfrac{1}{2}} 
= \frac{1}{2}\begin{bmatrix*}[r] 1+i & 1+i \\ -1-i & 1+i \end{bmatrix*}, ~
T= \begin{bmatrix} 1 & 0 \\ 0 & e^{\pi i / 4} \end{bmatrix},
\]\\
 Here $X^{\sfrac{1}{2}}=HPH$ and $Y^{\sfrac{1}{2}}=HZ$, but $T$ gates cannot be expressed this way. Adding $T$ gates hampers polynomial-time simulation~\cite{AG_2004}\footnote{The runtime of {\em our} simulation algorithms is not undermined by $T$ gates.} and enables universal quantum computation \cite{ekert2000basic,shi2002toffoli}.
 Unlike generic gates, this gate library supports quantum error correction \cite{NC}.
 
 Circuit depth is defined as maximum length of a monotonic path of unitary gates through the circuit.
% \begin{example}
 Figure \ref{fig:reorder} shows two equivalent circuits of depth 1+4+1 (1+ and +1 represent the initial and final rounds of Hadamards).
 %\end{example}

\begin{figure}[b]
\vspace{-4mm}
\begin{lstlisting}[escapechar=@]
using idx_size = unsigned long long;

template<typename function>
void Apply1QGate(@\blue{cmplx}@* w, // wave function
    int num_qubits, int target, function& gate_func)
{
 const @\blue{idx\_size}@ w_size = 1ull << num_qubits,
 block_size = 1ull << (num_qubuts - target)
 num_iters_per_block = block_size / 2,
 gate_bitmask = (1ull << ((num_qubits - 1) - target)),
 offset_idx =  1ull << ((num_qubits - 1) - target);
 @\blue{idx\_size}@ idx[2] = {0, 0};
 
 while (block_idx < w_size) 
 {  if (block_idx != 0 && block_idx % (2 * block_size) == 0)
        block_idx += block_size
    if ((block_idx & gate_bitmask) == 0)
        { idx[0] = block_idx; idx[1] = offset_idx + block_idx;
        @\blue{cmplx}@* new_w[2] = gate_func(w[idx[0]], w[idx[1]]);
        w[idx[0]] = new_w[0]; w[idx[1]] = new_w[1];
        ++block_idx; }
    else block_idx += (block_idx & gate_bitmask); }
}
\end{lstlisting}
\vspace{-5mm}
\caption{\label{fig:1q} Simulating a one-qubit gate on a wave function.}
%\vspace{-3mm}
\end{figure}

\section{Baseline Simulation Algorithms}
\label{sec:baseline}
 
 Equation \ref{eq:CZ} illustrates how quantum circuits
 can be evaluated. First, order the gates left to right (parallel gates can be ordered arbitrarily), pad each gate with an identity matrix of an appropriate dimension via Kronecker products to 
 obtain a $2^n\times2^n$ matrix, and then multiply
 all those matrices in order. The resulting operator represents the entire circuit and can be multiplied by input state vectors to find output vectors. While mathematically simple, this method is enormously wasteful and usually infeasible in practice. Instead, one applies each gate to the state vector, to avoid matrix-matrix multiplications. A key insight
 in high-performance Schr\"odinger simulation is how {\em not to pad gates with identity matrices}~\cite{de_raedt_massively_2007,wecker2014liquid,de_raedt_massively_2018,PB,qHIPSTER,IBM2017breaking,Taihu,IBM2019storage}.
 
\begin{figure}[t]
    \vspace{-4mm}
    \centering
    \begin{tabular}{cc}
    \includegraphics[width=7.6cm]{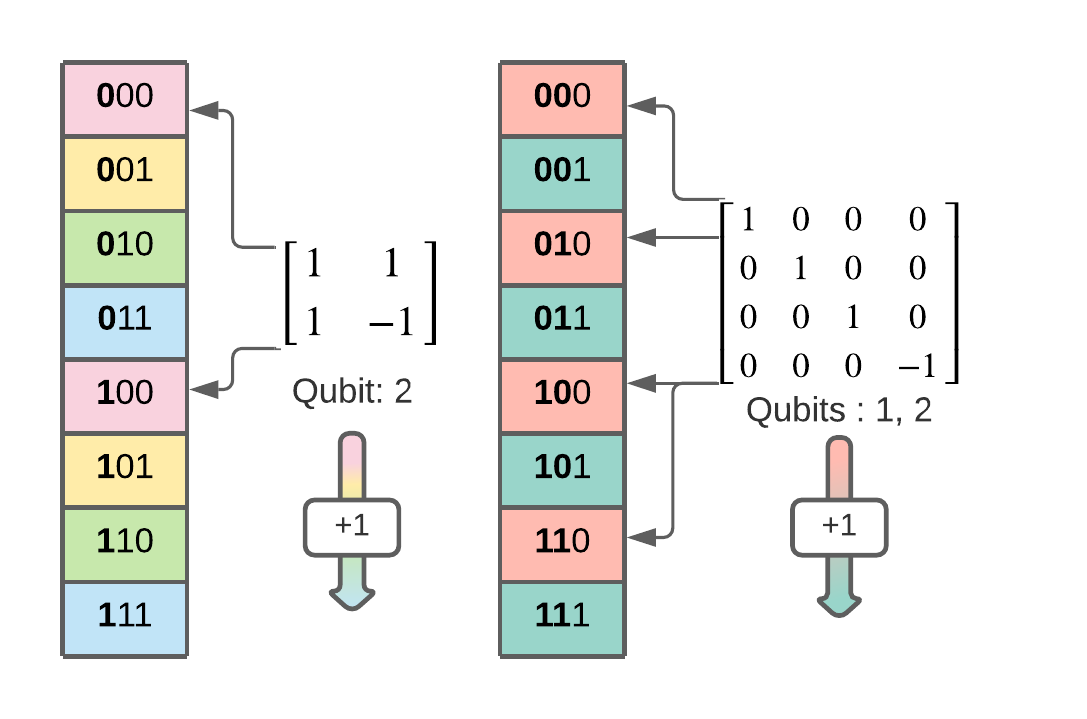}
    \end{tabular}
    \vspace{-7mm}
    \caption{
        \label{fig:baseline}
    Fast Schr\"odinger simulation of one- and two-qubit gates on a three-qubit state with 0-indexing (simplified). Consecutive address sets are shown in color.}
    \vspace{-3mm}
\end{figure}

 \noindent {\bf Fast Schr\"odinger simulation.}
 For a $q$-bit gate defined by its $2^q\times 2^q$ matrix and circuit qubits $i_0,\ldots,i_{q-1}$ to which it is applied, a typical simulation algorithm modifies the $2^n$-dimensional state-vector in-place. It traverses the state-vector and enumerates all $2^{n-q}$ disjoint sets of $2^q$ amplitudes, to which the gate should be applied {\em in-place}. To specify these sets,
 we turn to the binary ($n$-bit) representations of amplitude indices. Each set exhibits all $2^q$ combinations of bits indexed $i_0,\ldots,i_{q-1}$, whereas the remaining $n-q$ bits are common and form a {\em set id}.
 Each set can be produced by shifting the set containing $00\ldots0$
 by an appropriate amount. In general, the sets are not equally spaced, so linear loops illustrated in Figure \ref{fig:baseline} would not be sufficient. To set up efficient iteration, one finds (1) a minimal step size, (2) a maximum block size within which all sets are equally spaced, and
 (3) an update rule to find the next block. For generic gates, is common to implement such an iteration separately for 1-qubit and 2-qubit gates, based on the indices of qubits involved (Figure \ref{fig:1q}). However, some specific gates (e.g., diagonal) admit more efficient implementations. For gates that leave many index sets unchanged, such sets can be skipped.
 
%We illustrate fast Schr\"odinger simulation starting with 
\begin{example} $\mathit{CZ}$ gates are commonly used in circuit design and favored because they are qubit-symmetric and because
a $\mathit{CNOT}$ can be expressed via $\mathit{CZ}$ and $H$ gates. To simulate a single $\mathit{CZ}$ gate acting on qubits $i_0$ and $i_1$, note that it flips the sign of amplitude $\alpha_j$ when $j$ has 1s at binary positions $i_0$ and $i_1$, but otherwise leaves $\alpha_j$ unchanged.\footnote{This test can be implemented using bitmasks by first defining {\tt mask = 1ull << }$i_0$ { | 1ull << } $i_1$ and then checking {\tt  j \& mask == mask}.} Hence, the simulation traverses all amplitudes, and for each $\alpha_j$ decides whether to flip the sign based on the bits of $j$. Such simple {\em linear memory passes} are common for diagonal gates and benefit from standard CPU caching and prefetching policies.
\end{example}

Universal fault-tolerant quantum gate libraries often complement the $\mathit{CZ}$ gate with one-qubit gates \cite{ekert2000basic,shi2002toffoli}. Therefore, a minimal circuit-simulation framework can be completed with an algorithm to simulate an arbitrary one-qubit gate acting on qubit $i_0$ (simulating measurements is straightforward from the definition when all amplitudes are available). 
Diagonal one-qubit gates, such as $\mathit{Z}$ and $\mathit{T}$ can be simulated similarly to how we simulate $\mathit{CZ}$, but only one bit of the amplitude index $j$ is considered. 

\begin{example}
$\mathit{NOT}$ gates are not diagonal and swap pairs of amplitudes whose indices differ at bit $i_0$. One  simulates a $\mathit{NOT}$ gate in one pass over the state vector as follows: for each $j$, if bit $i_0$ is zero, swap $\alpha_j$ with $\alpha_{j'}$, where $j'$ differs from $j$ at bit $i_0$ only.
\end{example}

\begin{example} A generic one-qubit gate can be simulated by isolating the gate action to pairs of amplitudes whose indices differ in one bit only, such as \verb|46=b101110| and \verb|38=b100110|. 
Rather than swap these amplitudes (as for $\mathit{NOT}$ gates), it applies the $2\times 2$ gate matrix. Figure \ref{fig:1q} illustrates this with our C++ code, which uses bitwise instructions for efficiency. To scale this code beyond 64 qubits, our simulator redefines {\tt idx\_size}.
\end{example}
\begin{figure}[tb]
%\vspace{-8mm}
%\hspace{-7mm}
  \begin{lstlisting}[escapechar=@]
    /* Writes to idxs the 1st set of N=2^num_gate_qubits indices of amplitudes on which the gate can be applied.*/
    void GetFirstSetOfIndicesInWToApplyGate(
        int num_qubits // qubits in wave function, 
        @\blue{idx\_size}@* idxs, // starting idxs are written here
        @\blue{idx\_size}@ gate_bitmask)
    {
        const @\blue{idx\_size}@ num_gate_qubits =
        __builtin_popcountll(gate_bitmask);
        @\blue{idx\_size}@ num_idxs = 1, 
            stride = 1ull << (num_gate_qubits - 1);
        
        @\blue{idx\_size}@ prev_stride = stride;
        for (@\blue{idx\_size}@ i = idxs[0]; i < num_gate_qubits; ++i) {
            @\blue{idx\_size}@ q = __builtin_ctzl(gate_bitmask),
                shift = (num_circuit_qubits - 1) - q;
            for (@\blue{idx\_size}@ n = 0; num_idxs < (1ull << (i + 1)); n += prev_stride) {
                idxs[n + stride] = idxs[n] + (1ull << shift);
                ++num_idxs;
            }
            gate_bitmask ^= 1ull << q;
            prev_stride = stride;
            stride /= 2;
        }
    }
    \end{lstlisting} 
\vspace{-5mm}
\caption{\label{fig:extract} Our algorithm for extracting the first set of indices of the wave function to apply a generic $k$-qubit gate on qubits specified by the bitmask. Other sets of indices are obtained by shifting the first set, as shown in Figure \ref{fig:apply2gates}.
See Table \ref{tab:AVX} for compiler intrinsics. }
\vspace{-3mm}
\end{figure}

\begin{figure}[htb]
%\vspace{-4mm}
   \begin{lstlisting}[escapechar=@]
    template<typename function>
    void Apply2QGates(@\blue{cmplx}@* __restrict w, //wave function
        @\blue{idx\_size}@ gate_qubits, int num_qubits,
        function& gate_func, @\blue{idx\_size}@ collected_amps /* = 4 when gate_func is in AVX-256 */)
    {
        const @\blue{idx\_size}@ num_idxs = 4, w_size = 1ull << num_qubits,
        gate_bitmask = (1ull << ((num_qubits - 1) - __builtin_ctzl(gate_qubits))) |
            (1ull << ((num_qubits - 1) - (63 -  __builtin_clzl(gate_qubits))));
        
        array<@\blue{idx\_size}@, num_idxs> starting_idxs;
        /* Get starting indices into the wave function to start applying the gates on */
        GetFirstSetOfIndicesInWToApplyGate(num_qubits, starting_idxs.data(), gate_qubits);
        
        w = (@\blue{cmplx}@*)__builtin_assume_aligned(w, 64);
        @\blue{idx\_size}@ iters = 0, idx = 0; 
        array<@\blue{idx\_size}@, num_idxs> temp_idxs;
    
        const @\blue{idx\_size}@  num_iters = w_size / num_idxs;
        while (iters < num_iters) {
            /* Skip block where the gate has already been applied. This is where the bits in the wave function index are 1 at the position of the qubit value. */
            if (!(idx & gate_bitmask)) {
                iters += collected_amps;
                /* Increase the starting indices by idx to progress through the wave function */
                for (@\blue{idx\_size}@ i = 0; i < num_idxs; ++i)
                    temp_idxs[i] = starting_idxs[i] + idx;
                gate_func(w, temp_idxs.data());
                idx += collected_amps;
            }
            else idx += (idx & gate_bitmask);
        }
    }
    \end{lstlisting}
\vspace{-5mm}
\caption{\label{fig:apply2gates} Simulating a 2-qubit gate on a q. state using index extraction (Figure \ref{fig:extract}) and compiler intrinsics (Table \ref{tab:AVX}).
}
\vspace{-3mm}
\end{figure}

\section{Our Algorithmic Framework}
\label{sec:framework}
 We now introduce key techniques and optimizations
 for advanced Schr\"odinger-style simulation. First, we show
 how to achieve sufficient numerical accuracy with the 32-bit {\tt float} type
 and outline the benefits this brings. Second, we introduce a new gate-clustering
 approach that forms clusters of gates of a kind. Then we focus on optimizations for
 each gate type, and point out that clustering can be improved by circuit reordering.
 We use the gate library from Section \ref{sec:background}, but other common gates
 can be supported too.

%\vspace{-2.1mm}
\subsection{Data type selection and numerical accuracy}
\label{sec:float} 
  Given that state vectors consist of complex-valued amplitudes,
  we use a complex-valued type and need to choose a floating-point type to implement it. The two basic alternatives are the 32-bit {\tt float} and the 64-bit {\tt double}. Higher-precision types are also available and have been used for quantum simulation, but significantly increase computational load. Our choice of the 32-bit {\tt float} type improves memory footprint and throughput which happens to be a bottleneck for optimized simulation algorithms. Simulators that rely on the {\tt double} type are handicapped in memory and runtime (Section \ref{sec:empirical}).
  
  To facilitate the use of the {\tt float} type, we maintain numerical accuracy during simulation in the face of potential underflows. Among our gates, $\mathit{NOT}$, $\mathit{Z}$, $\mathit{CZ}$, and even $\mathit{T}$ gates do not significantly change the magnitudes
  of $\alpha_j$ values, but $\mathit{H}$, \Xsqr and \Ysqr include $1/\sqrt{2}$ or $1/2$ factors which,
  after hundreds of gates are applied, can lead to numerous underflows. To avoid underflows, we maintain a global power of $1/\sqrt{2}$ to accumulate contributions from individual gates.
  Specifically, we store a single integer value (starting with 0) and increment it whenever we encounter a factor of $1/\sqrt{2}$ (increment twice for $1/2$). This value is accounted for when reading off $\alpha_j$ values at the end of the simulation, but that sometimes leads to very large values. Therefore, we ``flush'' accumulated $(1/\sqrt{2})^p$ when $p>100$, back into $\alpha_j$. We use a similar counter $s$ for global phase $i^s$, but it cycles through only four possible values.
  By inspecting gate matrices in Section \ref{sec:background}, one can see that, after factoring out $1/\sqrt{2}$, all gates can be simulated without floating-point multiplies to improve speed and accuracy.
  
  \begin{example}
   To simulate a $T$ gate without floating-point multiplies, note that the only nontrivial multiplication involves $\exp(\pi i /4) = (i + 1)/\sqrt{2}$. This multiplication can be realized by first incrementing the $p$ count and then using $(i+1)z = iz + z =  (\mathrm{Re}(z) - \mathrm{Im}(z)) + i(\mathrm{Re}(z) + \mathrm{Im}(z))$. In other words, add a complex number $z$ to its product by $i$, the latter computed by swapping the real and imaginary parts and negating the real part. Also see Example \ref{ex:AVX}.
 \end{example}
  As per Section \ref{sec:cluster}, our simulator rarely deals with $T$ gates one by one, but rather clusters them and
  simulates entire clusters, eliminating not only floating-point multiplies, but also most floating-point additions and subtractions by using integer arithmetic and bit-parallel instructions instead.

  When relying on the compact {\tt float} type, it is important to explicitly check for accuracy loss. Since quantum states are represented by norm-one vectors, we compute the norm
  before measurement and check how close it is to 1. A careless norm computation can introduce greater errors than our simulation when small contributions of individual amplitudes $\alpha_j$ are accumulated
  in a much larger running sum. Indeed, small contributions of individual amplitudes $\alpha_j$ are accumulated in a much larger running sum. Adding a tiny number to a much larger number exposes mantissa limitations. This problem can be mitigated by representing the running sum by the higher-precision {\tt double} type and/or by using a redundant sum-of-two-values representation, a common numerical technique for robust arithmetics~\cite{Ogita_2005}. The norm computation can be additionally optimized using methods of Section \ref{sec:HW}. In particular,
  using compact {\tt float} values to represent amplitudes $\alpha = \mathrm{Re} + i~\mathrm{Im}$ offers an additional benefit: four pairs of complex numbers can be quickly read from and written to memory, multiplied and added using AVX-2 instruction.
  
\subsection{Gate clustering and bitmask encoding}
\label{sec:cluster}

Simulating gates one at a time is slow because it requires separate memory traversals. Therefore, it is common to simulate gate clusters in batches. We develop a technique that clusters more gates, respects our avoidance of floating-point multiplies, and enables efficient representations with bitmasks and downstream optimizations.

Prior quantum simulators typically cluster adjacent gates acting on the same qubits (when this is possible). Google QSim merges each one-qubit gate to some nearby two-qubit gate, whereas the simulator in 
\cite{PB} clusters gates up to five qubits, multiplies out gate matrices up to $32\times 32$, and optimizes matrix-vector products with SIMD multiply-accumulate instructions \cite{PB}. These clusters are $O(1)$ in size, and their matrices lose the structure that helps us eliminate most floating-point multiplies.  In contrast, our approach
   \begin{itemize}
   \item creates clusters that grow as $O(q^2)$ with $q$ qubits,\footnote{
   In Google supremacy benchmarks, we find clusters with $O(q)$ CZ gates,
   but circuits with larger clusters can be shown.
   }
   \item considerably reduces memory traversals,
   \item avoids floating-point multiplies and MAC instructions. 
   \end{itemize}
The main insight is to cluster adjacent gates of a kind --- diagonal gates ($\mathit{T}$ and $\mathit{CZ}$) separately from one-qubit non-diagonal gates ($H$, \Xsqr, \Ysqr) as illustrated in Figure \ref{fig:reorder}.
%This clustering enables bitmask encodings of constituent gates.
% Clustering diagonal gates together and one-qubit gates together brings a number of benefits. 
   In simulating diagonal gates, we rely on the fact that
   they act on individual $\alpha_j$ amplitudes without permuting or mixing them. In particular, the order in which diagonal gates are applied (within the cluster) does not matter. Nor does the order of one-qubit gates acting on different qubits.\footnote{Clustering gates of a kind helps find hidden gate cancellations. Such gate cancellations sometimes appear in compiled circuits, but not in well-designed simulation benchmarks we use \cite{GoogleSuprem,new_benchmarks}. Thus, our experiments showcase other benefits of such clustering.} For each type of one-qubit gate ($T$, etc), we encode circuit gates in each cluster using bitmasks.
   The bitmasks power downstream optimizations with profound impact on simulation performance, as we show later.
   
   \begin{example}
   In Figure \ref{fig:reorder}, the four-qubit circuit on the right contains a cluster of $T$ gates on qubits 0-3. This cluster can be
   represented by the bitmask \verb|15=b1111|, neglecting the order in which these gates were listed in the circuit. The same encoding is used for $X^{\sfrac{1}{2}}$ gates (\verb|7=b0111|) and $Y^{\sfrac{1}{2}}$ gates (\verb|14=b1110|).
   \end{example}
 
   A single bitmask cannot encode multiple $T$ gates on one qubit, but such gates can be separated into adjacent layers (cycles) and captured using one bitmask per layer.\footnote{ The benchmarks used in this work \cite{GoogleSuprem,new_benchmarks} do not include repeated gates.}
   Bitmask encodings of $\mathit{CZ}$ gates are more involved and discussed in Section \ref{sec:diag}.
   
   So far, we explained which gates we cluster and outlined
   the logic behind the approach. As will be seen in Section \ref{sec:diag}, our use of bitmasks significantly reduces the number of floating-point operations by ($i$) consolidating the phases contributed by $\mathit{CZ}$ and $T$ gates to each amplitude index $j$,
   and ($ii$) applying the resulting phases to the amplitudes $\alpha_j$, when the phases are $\neq 1$.
   Additionally, simulating all diagonal gates in one memory pass over amplitudes $\alpha_j$ reduces memory traffic that is often the main limiting factor when large amounts of memory 
   are used.
   
   Optimizations for non-diagonal gates are covered in Section
   \ref{sec:nondiag}. For algorithmic details on gate clustering
   see Section \ref{sec:reorder}.
   
\subsection{Optimizations for diagonal gates}
\label{sec:diag}

   Our use of gate clustering before simulation (Section \ref{sec:cluster}) creates opportunities to optimize algorithms for each gate type and
   develop batched simulation. Here we focus on diagonal gates and clusters,
   whose distinctive property is that they modify individual amplitudes 
   $\alpha_j$ without mixing or permuting them. After discussing their prevalence in quantum circuits, we outline the handling of generic diagonal gates and then focus on diagonal gates that are commonly used
   and available on recent Google chips. Here we gain efficiency through
   bit-parallel operations.

   Clusters of diagonal gates appear in quantum combinatorial algorithms (Shor's and Grover's) and Hamiltonian simulations \cite{Bullock_Markov_2004,NC}.
   A particularly ubiquitous example is the Quantum Fourier Transform implemented with controlled $R_z$ rotations that finds uses in arithmetic circuits \cite{Ruiz_Perez_2017}, phase estimation, and Hamiltonian simulation. Diagonal clusters can be simulated by traversing the entire state vector only once if contributions of all gates in the cluster are aggregated for each amplitude $\alpha_j$.
   When a cluster spans $<20$ qubits, the gates can be multiplied out into an array, and each array element multiplies bit-compatible amplitudes. Below, we optimize the simulation of $\mathit{T}$ and $\mathit{CZ}$ gates.
   
   For each gate in the cluster, the task is simple. For example, a $\mathit{T}$ gate acting on qubit $i_0$ leaves unchanged those
   $\alpha_j$ values where the binary form of $j$ has 0 at bit position $i_0$, and multiplies
   the remaining $\alpha_j$ by $\exp(\pi i/4)$. A 
   $\mathit{CZ}$ gate acting on qubits $i_0$ and $i_1$
   negates $\alpha_j$ when $j$ has bits $1$ at positions $i_0$ and $i_1$.
   
   \begin{figure}[tb]
   {\small
        \begin{quantikz}
        0 \ket{0} &  \qw  & \qw \\
        1 \ket{0} & \ctrl{1}  & \qw \\
        2 \ket{0} & \gateZ  & \qw \\
       \end{quantikz} \hspace{0.3cm} $\longmapsto$
    }
       \hspace{0.3cm}
    {
    \small
       \begin{tabular}{|c|} \hline
        CZ\_bitmasks \\ \hline
        0 : 000 (skip)  \\ \hline
        1 : 100 \hspace{0.7cm} \\ \hline
        2 : 010 \hspace{0.7cm} \\ \hline
       \end{tabular}
     } \\
     \vspace{-4mm}
        \includegraphics[width=9cm]{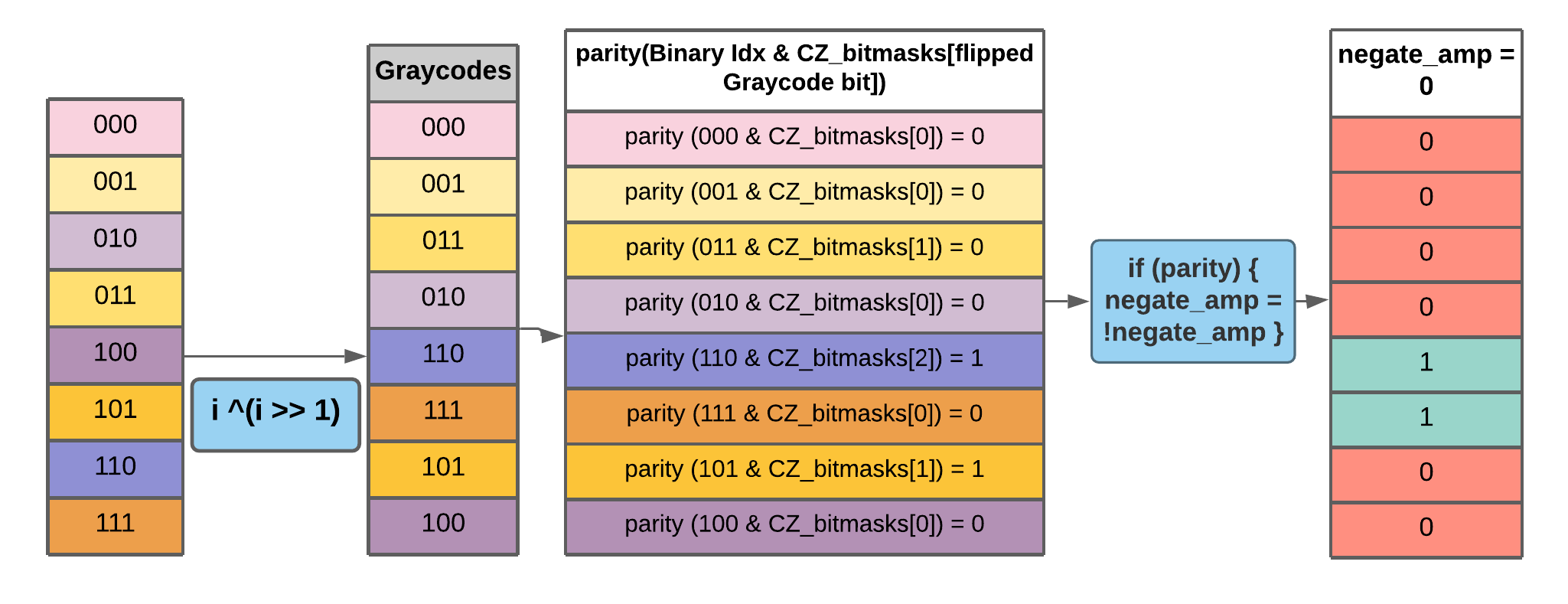}
     \vspace{-6mm}
       \caption{ \label{fig:CZ}
       Encoding CZ gates on three qubits using bitmasks and simulating them in one pass using Gray codes. For each address, we decide whether the respective amplitude should be negated. Gray codes support incremental computation 
       over adjacent addresses based on
       which bit flips.
       }
   \vspace{-4mm}
   \end{figure}
   
   \begin{figure}[htb]
%\vspace{-5mm}
    \begin{lstlisting}[escapechar=@]
    /* Returns the 8 phases effected by a set of CZ gates on 8 consecutive amplitudes */
    unsigned char GetCZPhaseInBlockUsingGrayCodesAndBitmask(
        @\blue{idx\_size}@ num_qubits,
        // one bitmask per qubit
        @\blue{idx\_size}@* __restrict CZ_bitmasks, 
        @\blue{idx\_size}@ block_idx /* block of 8 floats */)
    {
        // gc : gray codes
        @\blue{idx\_size}@ prev_gc = (block_idx - 1) ^ ((block_idx - 1) >> 1), gc0 = prev_gc, gc4 = (block_idx + 4) ^ ((block_idx + 4) >> 1);
        const @\blue{idx\_size}@ gc[8] = {gc0, gc0 ^ 1, gc0 ^ 3, gc0 ^ 2, gc4, gc4 ^ 1, gc4 ^ 3, gc4 ^ 2};
        unsigned char z_phase_result = 0u;
        /* In blocks of 4, indices 0,1,0 capture the application of CZ on the least-sig  qubit. */
        const @\blue{idx\_size}@ bit_idx[8] = {__builtin_ctzl(gc[0] ^ prev_gc), 0, 1 , 0, __builtin_ctzl(gc[4] ^ gc[3]), 0 , 1, 0};
        // Get CZ parity for the prev. set of 1-bit indices
        @\blue{idx\_size}@ gate_count = 0;
        for (@\blue{idx\_size}@ i = 0; i < num_qubits; ++i) 
           if (((prev_gc & (1ull << i)) == (1ull << i))
                && (prev_gc & CZ_bitmasks[i]))  
                gate_count += __builtin_popcountll((prev_gc & CZ_bitmasks[i]));
        if (gate_count & 2) z_phase_result |= 1 ;
        /* Update CZ gate state by calculating 
           the new parity in the current block */
        for (int i = 0; i < 8; ++i) 
            if (__builtin_parityl(CZ_bitmasks[bit_idx[i]] & gc[i]) == 1)
                z_phase_result |= z_phase_result & (1 << i) ? 0: 1 << i;
        return z_phase_result
    }
    \end{lstlisting}
    \vspace{-5mm}
    \caption{
    \label{fig:Gray}
    Our optimized algorithm for simulating a bitmask-encoded cluster of $CZ$ gates on a block of consecutive amplitudes. It performs a loop-unrolled Gray-code traversal on a block of 8 consecutive indices. The 8 returned phase values determine whether or not to negate each amplitude in the wave function. Compiler intrinsics are explained in Table \ref{tab:AVX}.
    }
\vspace{-2mm}
\end{figure} 
   
   The handling of diagonal clusters can be optimized further. We form $n$-qubit layers of $\mathit{T}$ gates, where each layer has at most one gate on any given qubit and can thus be encoded by a bitmask $m$, where each gate location is represented by a 1 bit. Bitmasks are formed before the memory pass. For each amplitude $\alpha_j$, each bit of each bitmask may contribute a factor of $\exp(\pi i /4)$ or a factor
   of 1. The nontrivial contribution occurs when a 1-bit in bitmask $m$ matches a 1-bit in index
   $j$.
   
   \begin{example}
   To count pairs of matching 1-bits between an amplitude index $j$ and a $T$-gate bitmask $m$, two single-cycle CPU instructions suffice: {\tt popcount(m \& j)}. See Table \ref{tab:AVX} for more details. 
   \end{example}

   Since $T^8$=I, the number of matching bits above can be taken $\mathrm{mod}$-8. Applied as a power to $\exp(\pi i /4)$, this integer can give eight values:  $\pm 1$, $\pm i$ and $(\pm 1 \pm i)/\sqrt{2}$. To multiply by these values, we use increments of the $p$ counter,
   floating-point negations, swaps of real and imaginary parts, and additions (Section \ref{sec:float}).

   To leverage fast bit-based CPU instructions, bitmasks are stored in 64-bit integers when 
   simulating $\leq 64$ qubits. Processing dozens of $T$ gates in a cluster by several bit-based operations per amplitude is much more efficient than simulating gates one by one.
   
   For simulation, each $\mathit{CZ}$ gate can be encoded by a bitmask $m$ with two nonzeros, then such bitmasks are stored in a list (as long as the number of $\mathit{CZ}$ gates). For each $\alpha_j$ and each $\mathit{CZ}$ gate, we can check if {\tt m \& j == m} bitwise, in which case we increment a counter of contributions. Since each $\mathit{CZ}$ gate contributes only a $\pm1$ factor, contributions are aggregated and then we either apply the resulting -1 or do nothing (saving a memory write).
   
   When dealing with large clusters of $\mathit{CZ}$ gates on $n$ qubits, a more efficient approach is
   to use (up to $n$) bitmasks that capture multiple gates each. For qubit $k\geq 0$, the bitmask $m_k$ represents (qubits $l\neq k$ of) $\mathit{CZ}$ gates that also act on qubit $k$. To each $\alpha_j$, these gates can cumulatively contribute phase 1 or -1, which we determine by aggregating \verb|parityll(|$m_k$ \verb|& j)| over all $k$ such that \verb|j & (1ull << k)| $\neq 0$.
   
   \begin{example}
   Consider a cluster of six $\mathit{CZ}$ gates that couple all pairs of four qubits. This cluster is encoded by the following
   set of bitmasks (one per qubit):
   $m_0=$\verb|1110|
   $m_1=$\verb|1101|, $m_2=$\verb|1011|, $m_3=$\verb|0111|. Note that there are exactly twelve nonzero bits total across these bitmasks.
   \end{example}
   
   Further optimizations use Gray codes. Specifically, we traverse $\alpha_j$ in a Gray code order, so that $j$ changes one bit at a time to minimize necessary updates.
   In this work, we use the more-common {\em reflected} Gray code
   that can be produced from a regular counting sequence $k=0,1,2,3,\dots$ with bit operations {\tt j = k $\oplus$ ( k >> 1 )}.
   
   \begin{example} The three-bit {\em reflected Gray code} uses codewords
   \verb|000-001-011-010-110-111-101-100| or 0-1-3-2-6-7-5-4. Note that this code is cyclic --- the first and the last values differ in one bit. It can be obtained by first reflecting the first half and then by setting the most significant bit to 1.
   \end{example}
   
    Given a pattern of $\mathit{CZ}$ gates in a bitmask-encoded cluster, we precompute which $\mathit{CZ}$ gates become active (-1) or inactive (1) when each bit switches. When processing blocks of indices,
   instead of index calculations from scratch, we incrementally update the ``state'' from the previous block. This technique, shown in Figure \ref{fig:CZ}, reduces the complexity of amplitude updates from $O(n)$ to $O(1)$ time, after initialization.
   
   Figure \ref{fig:Gray} implements the ideas above, using advanced CPU instructions via compiler intrinsics
   (Table \ref{tab:AVX}). The code works with bitmask-encoded $\mathit{CZ}$ gates and finds the implied $\mathit{Z}$ phase changes for a block of 8 amplitude indices. Returned as a byte, these 8 bits determine if respective amplitudes must be negated. The function can be used in a thread-parallel traversal of the wave function in conjunction with aligned memory reads (Section \ref{sec:HW}).

Other common diagonal gates can be simulated natively or by expressing them
via supported gates, e.g., $P=T^2$ and $Z=T^4$.

\subsection{Optimizations for non-diagonal gates}
\label{sec:nondiag}
  
  We start with gate-specific optimizations to reduce computation, and then in Section \ref{sec:HW} present more general optimizations that reduce memory accesses and work with arbitrary gates.
  Recall that all non-diagonal gates in our gate library are one-qubit gates. If one wanted to simulate a $\mathit{CNOT}$ gate, it can be re-expressed using a $\mathit{CZ}$ gate and two $\mathit{H}$ gates on the sides.
  Thus, we are now simulating the following gates (leading factors extracted):
\[
H' = \begin{bmatrix*}[r] 1 & 1 \\ 1 & -1 \end{bmatrix*}, ~
X^{\sfrac{1}{2}'} = \begin{bmatrix*}[r] 1+i & 1-i \\ 1-i & 1+i \end{bmatrix*}, ~
Y^{\sfrac{1}{2}'} = \begin{bmatrix*}[r] 1+i & 1+i \\ -1-i & 1+i \end{bmatrix*}
\]
   When different non-diagonal gate types are applied on the same qubit, their order matters.
   However, gates applied on different qubits can be reordered. Thus, we cluster
   gates of each kind into layers, and represent each layer by a bitmask $m$. Since
   applying gates one at a time is inefficient, we apply them two at a time. While this
   requires fetching more data at a time, the resulting matrices
   \hush{--- 
   $H\otimes H$, $X^{\sfrac{1}{2}}\otimes X^{\sfrac{1}{2}}$,
   $Y^{\sfrac{1}{2}}\otimes Y^{\sfrac{1}{2}}$.}
   \[
H'\otimes H' = \begin{bmatrix*}[r] 1 & 1 & 1 & 1 \\ 1 & -1 & 1 & -1 \\ 1 & 1 & -1 & -1 \\ 1 & -1 & -1 & 1 \end{bmatrix*}, Y^{\sfrac{1}{2}'} \otimes Y^{\sfrac{1}{2}'} = i \begin{bmatrix*}[r]  1 & 1 & 1 & 1 \\ -1 & 1 & - 1 & 1 \\ -1 & -1 & 1 & 1 \\ 1 & -1 & -1 & 1 \end{bmatrix*},
\]
\[
X^{\sfrac{1}{2}'}\otimes X^{\sfrac{1}{2}'} = \begin{bmatrix*}[r] i & 1 & 1 & - i \\ 1 & i & - i & 1 \\ 1 & - i & i & 1 \\ - i & 1 & 1 & i  \end{bmatrix*}
\]

 use  mostly $\pm 1$ and $\pm i$ as their entries and can be multiplied by without floating-point multiply and MAC instructions. For example, multiplying a complex value $\alpha_k$ by the imaginary $i$ entails a swap of the real and imaginary parts and one negation. Using such observations, we have developed several algorithms in the spirit of Figure \ref{fig:1q}.
   A common pattern in these algorithms is that the two-qubit gate combination acts each time on {\em four} amplitudes whose indices differ by two bits. To extract such indices,
   we find the first set using the code
   in Figure \ref{fig:extract} and obtain the remaining sets by shifting
   the indices as shown in Figure \ref{fig:apply2gates}.
   
   Our first implementation uses a generic
   \verb|gate_func| as in Figure \ref{fig:1q}, but is not limited
   to one-qubit gates. As shown in
   Figure \ref{fig:apply2gates}, it extracts sets of amplitude indices,
   then for each set loads the amplitudes from the wave function, applies \verb|gate_func| to
   them and saves the result back into the wave function. Our second implementation in Section \ref{sec:AVX} uses the same index-extraction mechanism, but increases data-level parallelism via custom code for each pair of one-qubit gates (hence, no generic gate function is passed). 
   A small number of unpaired one-qubit gates are simulated individually.

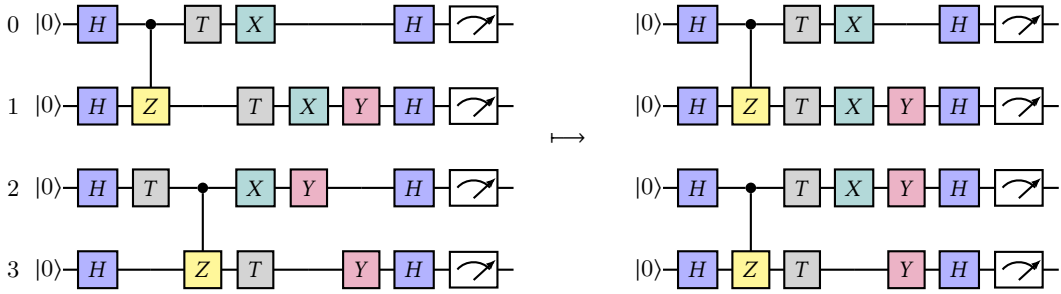
\begin{figure*}[ht]
    \begin{quantikz}[column sep=0.2cm]
        0 & \ket{0}& \gateH & \ctrl{1} & \gateT & \gateX & \qw & \qw & \gateH & \meter{} & \qw \\
        1 & \ket{0} & \gateH & \gateZ &  \qw & \gateT & \gateX & \gateY  & \gateH &  \meter{} & \qw \\
        2 & \ket{0} & \gateH & \gateT & \ctrl{1} & \gateX & \gateY & \qw & \gateH &  \meter{} & \qw \\
        3 & \ket{0} & \gateH & \qw & \gateZ & \gateT & \qw & \gateY & \gateH &  \meter{} & \qw
    \end{quantikz}
    \hspace{0.3cm} $\longmapsto$ \hspace{0.3cm}
    \begin{quantikz}[column sep=0.2cm]
        \ket{0} & \gateH & \ctrl{1} & \gateT & \gateX & \qw & \gateH &  \meter{} & \qw \\
        \ket{0} & \gateH & \gateZ  & \gateT & \gateX & \gateY & \gateH &  \meter{} & \qw \\
        \ket{0} & \gateH & \ctrl{1} & \gateT & \gateX & \gateY & \gateH &  \meter{} & \qw \\
        \ket{0} & \gateH & \gateZ & \gateT & \qw & \gateY & \gateH &  \meter{} & \qw
    \end{quantikz}
%\vspace{-4mm}
\caption{\label{fig:reorder} 
Clustering gates of a kind by reordering, with the algorithm from Section \ref{sec:reorder}. In the figure, gates are ordered top down and left to right, to make clusters contiguous. The $X$ and $Y$ boxes above represent  \Xsqr and \Ysqr gates, respectively. Circuit depth is 1+4+1, and the circuit (not including the initial and final Hadamard gates) can be encoded using the following bitmasks as follows. $\mathit{CZ}$ gates: $m_1$={\tt b0001}, $m_3$={\tt b0100}; $T$ gates: {\tt b1111};
\Xsqr gates: {\tt b0111}; \Ysqr gates: {\tt b1110}.}
\end{figure*}

\subsection{Gate clustering by reordering}
\label{sec:reorder}
   To perform clustering outlined in Section \ref{sec:cluster},
   we assume a quantum circuit specified by a list of gates (Section \ref{sec:background}) ordered so that every gate $g_a$ whose output qubit acts
   as input to gate $g_b$ appears before $g_b$ 
   (parallel gates are ordered arbitrarily). Such {\em topologically sorted} orders are generally not unique. Moreover, diagonal gates can always be reordered without affecting circuit functionality.
   \hush{However, this representation does not expose adjacency between gates acting on the same qubits, which is key to clustering. In particular, two adjacent one-qubit gates may be listed with another gate in between that acts on a different qubit.} Google benchmarks
   \cite{GoogleSuprem,new_benchmarks} additionally pack parallel gates into numbered ``cycles'', but we ignore this additional structure.
   \hush{
   Instead, we use the data structure from \cite{Prasad_2006}
   which overlays adjacency edges on top of a list of gates.
   }

   When gates of a kind are adjacent in the given gate ordering,
   they form a natural cluster. However, non-adjacent gates of a kind that can be reordered to form larger clusters.
   The search proceeds from the beginning of the circuit in a topological order. Start with a gate $g_k$ that cannot be in a cluster formed before (such as the first gate in the circuit) and assume that no gates after it can be in a cluster formed before (or else they would have been reordered). The inner loop of the algorithm (for a fixed $g_k$) finds gates that cluster with $g_k$ and reorders them accordingly. The outer loop goes over all yet-unclustered gates $g_k$. The state of the inner loop designates each qubit as {\em unobstructed} (initially) or {\em obstructed}. An {\em obstructed} qubit prevents a same-type gates from being moved next to $g_k$.
   
   The inner loop scans (traverses) gates after $g_k$, classifies each gate $g_l$ in one of three categories and performs the following actions:

   \begin{enumerate}
%      \vspace{-2mm}
      \item a gate of the same kind as $g_k$ --- if all of the gate's qubits are {\em unobstructed}, then reorder the gate toward $g_k$ to form a larger cluster, else mark all of the gate's qubits {\em obstructed};
      \hush{\footnote{The {\em else} clause is unnecessarily pessimistic and can be refined.}}
      \item a gate $g_l$ of a different kind that can be reordered with $g_k$ --- do not change qubit designations because gates of the same kind as $g_k$ can be reordered past $g_l$ towards $g_k$;
      \item a gate that cannot be reordered with a $g_k$ --- mark all of the gate's qubits as {\em obstructed}.
   \end{enumerate}
   The scan from $g_k$ continues until all qubits are marked as {\em obstructed} or all gates have been scanned. Upon the completion of the scan, all qubits are reset to {\em unobstructed}, and the next iteration of the outer loop focuses
   on to the next gate not in a cluster.
   Each gate is guaranteed to be in a cluster (single-gate clusters are allowed).
   
   \begin{example}
     The circuit in Figure \ref{fig:reorder} starts and ends with a cluster of Hadamards, unchanged during reordering.\footnote{Starting Hadamards are applied to the initial state $\ket{0\ldots 0}$, so can be replaced by initializing the state into a full superposition.} When the outer loop focuses on the $\mathit{CZ}$ gate on qubits 0-1, it reorders the other $\mathit{CZ}$ gate to be adjacent. Here we recall that $\mathit{CZ}$ and $T$ gates are diagonal and so can always be swapped.
     The second iteration clusters $T$ gates. Subsequent iterations cluster
     \Xsqr and \Ysqr gates.
   \end{example}
   
   The example in Figure \ref{fig:reorder} suggests additional optimizations for adjacent clusters of one-qubit gates. In particular, separate 
   memory passes for a \Ysqr$\otimes$\Ysqr pair and a $H \otimes H$ pair acting on the same qubits can be coalesced into a single pass as follows.
   \begin{equation}
    (Y^{\sfrac{1}{2}} \otimes Y^{\sfrac{1}{2}}) (H \otimes H) =
    (Y^{\sfrac{1}{2}} H \otimes Y^{\sfrac{1}{2}} H)
   \end{equation}
   Such coalesced matrices inherit simple structure from Kronecker-product matrices
   shown in Section \ref{sec:nondiag}.
   Moreover, $Y^{\sfrac{1}{2}} H \otimes Y^{\sfrac{1}{2}} H$ is diagonal, which further simplifies simulation. 
   One can also merge unpaired one-qubit gates, e.g., the \Xsqr and $H$ gates at the top qubit line in Figure 
   \ref{fig:reorder}. These efficiency improvements require more dedicated simulation kernels like the one illustrated in Figures \ref{fig:XX} and \ref{fig:XXAVX}.
   In Section \ref{sec:AVX}, we introduce the use of 
   aligned memory read-write instructions, which make it possible to apply these kernels on entire L1 cache lines. 
   This effectively clusters one-qubit gates so that 
   they can be applied on the same L1 cache line.

   Clustering by reordering is performed once, before simulation starts. To get larger clusters,
   we note that multiplying by
   $X^{\sfrac{1}{2}}\otimes Y^{\sfrac{1}{2}}$ 
   is as simple as for other
   Kronecker products in Section
   \ref{sec:nondiag}. Thus, we blend $X^{\sfrac{1}{2}}$ and $Y^{\sfrac{1}{2}}$ clusters. This circuit preprocessing produces a {\em simulation plan} that
   defines and schedules individual gate sets
   \hush{(clusters of diagonal gates, paired one-qubit gates, coalesced pairs and unpaired one-qubit gates, etc)} 
   handled directly by our algorithms. Then this plan is executed. Our work leaves significant room for simulation plan optimization.

 \begin{figure}[!t]
%\vspace{-1mm}
    \begin{lstlisting}[escapechar=@]
//-0.0f is needed to switch real and imaginary signs in the xor operations due to multiplication by - i .
const @\blue{\_\_m256}@ kneg1 = {-0.0f, 0.0f, -0.0f, 0.0f, -0.0f, 0.0f, -0.0f, 0.0f};

void ApplyXX12GateAVX(@\blue{cmplx}@* __restrict w, // wave function
                      @\blue{idx\_size}@ target[4])
{
    // temp wave function
    float* __restrict t_w = (float*)__builtin_assume_aligned(w, 64); 
    
    @\blue{\_\_m256}@ a0 = _mm256_load_ps (t_w + 2*target[0]), 
    a1 = _mm256_load_ps (t_w + 2*target[1]),
    a2 = _mm256_load_ps (t_w + 2*target[2]),
    a3 = _mm256_load_ps (t_w + 2*target[3]);

    const @\blue{\_\_m256}@ t0 = _mm256_add_ps(a0, a3);
    const @\blue{\_\_m256}@ t1 = _mm256_add_ps(a1, a2);
    @\blue{\_\_m256}@ t2 = _mm256_sub_ps(a0, a3);
    @\blue{\_\_m256}@ t3 = _mm256_sub_ps(a1, a2);
    // Permute real and imaginary numbers
    t2 = _mm256_permute_ps(t2, 0b10110001);
    t2 = _mm256_xor_ps(t2, kneg1);
    t3 = _mm256_permute_ps(t3, 0b10110001);
    t3 = _mm256_xor_ps(t3, kneg1);

    a0 = _mm256_add_ps(t1, t2); a1 = _mm256_add_ps(t0, t3);
    a2 = _mm256_sub_ps(t0, t3); a3 = _mm256_sub_ps(t1, t2);
    
    _mm256_store_ps(t_w + 2*target[0], a0);
    _mm256_store_ps(t_w + 2*target[1], a1);
    _mm256_store_ps(t_w + 2*target[2], a2);
    _mm256_store_ps(t_w + 2*target[3], a3);
}
    \end{lstlisting}
\vspace{-5mm}
    \caption{\label{fig:XXAVX} Optimized AVX-2 code to apply the $X^{\sfrac{1}{2}} \otimes X^{\sfrac{1}{2}}$ gate on four amplitudes loaded onto CPU registers. Code for the $Y^{\sfrac{1}{2}} \otimes Y^{\sfrac{1}{2}}$ gate is simpler due to its simpler matrix, as seen in Section \ref{sec:nondiag}. Compiler intrinsics are explained in Table \ref{tab:AVX}.}
\vspace{-6mm}
\end{figure}

\begin{table}[!tb]
%\vspace{-2mm}
\begin{center}
\begin{tabular}{|p{2.5cm}||l||p{3cm}|}
\hline
\sc Instruction & \sc Compiler & \sc Use in \\ 
\sc type & \sc intrinsics & \sc simulation\\
\hline
\centering
Aligned read/write &
\begin{lstlisting}
mm256_load_ps
mm256_store_ps
\end{lstlisting} & 
\vspace{-5mm}
For all gates, loads/stores amplitudes between RAM \& registers.\\

\hline
\centering
Packed 32-bit {\tt float} arithmetics &
\begin{lstlisting}[escapechar=@]
mm256_add_ps
mm256_sub_ps
mm256_mul_ps
mm256_fmadd_ps
mm256_fmsub_ps
\end{lstlisting} & 
\vspace{-9mm}
Can be used with generic simulation of arbitrary gates. In RR, used with \Xsqr, \Ysqr, and $H$ gates. AVX multiplication is optionally used with $T$ gates. \\
\hline
\centering
Bitwise ops on packed 32-bit {\tt float}s  &
\begin{lstlisting}
mm256_xor_ps
mm256_or_ps
\end{lstlisting} 
% mm256_and_ps
& With $X^{\sfrac{1}{2}}$, $Y^{\sfrac{1}{2}}$, $H$ gates.
\hush{Bitwise AND is used in the kernel for decomposed $CZ$ gates in the Feynman simulation.}\\

\hush{
\hline
\centering
Packed 32-bit {\tt float}s mask and compare &
\begin{lstlisting}
mm256_movemask_ps
mm256_cmp_ps
\end{lstlisting} & Used together to compare registers in the kernel for decomposed $CZ$ gates in the Feynman simulation.\\
}

\hline
\centering
Packed 32-bit {\tt float}s swizzle &
\begin{lstlisting}
mm256_shuffle_ps
mm256_permute_ps
\end{lstlisting} &
\vspace{-3mm}
Used to rearrange re and im parts of complex amplitudes for arithmetic optimizations.\\

\hline
\centering
Assume aligned &
\begin{lstlisting}
builtin_
assume_aligned
\end{lstlisting} &
%\vspace{-1mm}
In most kernels lets the compiler optimize for aligned vectors.\\

\hline
\centering
Count trailing, leading 0-bits &
\begin{lstlisting}
builtin_ctzl
builtin_clzl
\end{lstlisting} & 
%\vspace{-1mm}
Used to extract indices and apply bitmasks.Can be used to apply arbitrary 1q gates in parallel using bitmasks.\\

\hline
\centering
Count 1-bits &
\begin{lstlisting}
builtin_ 
popcountll
\end{lstlisting} & Used with $CZ$ and $T$ gates, also to extract data from bitmasks.\\

\hline
\centering
Parity of 1-bit count &
\begin{lstlisting}
builtin_
parityll
\end{lstlisting} & Used with $Z$ and $CZ$ gates, with Gray codes.
%Used in the kernel for decomposed $CZ$ gates for Feynman simulations.
\\
\hline

\end{tabular}
%\parbox{8cm}
{
\caption{\label{tab:AVX} Compiler intrinsics used to accelerate simulation.
}
}
\end{center}
\vspace{-4mm}
\end{table}

\section{Leveraging the CPU Architecture}
\label{sec:HW}
   
   Section \ref{sec:framework} dramatically reduces computation versus the baseline Schr\"odinger framework in Section \ref{sec:baseline} and also
   reduces memory accesses for diagonal gates
   to one linear pass per cluster.
   New efficiencies for non-diagonal gates are unlocked by enhancing memory locality, optimizing algorithms for the CPU cache size (``blocking''), leveraging data-level parallelism, and using multiple CPU threads. 
   Section \ref{sec:ablation} shows that these optimizations greatly
   improve performance. Most of them can be used with arbitrary one-qubit gates.
   
\subsection{Memory locality and CPU cache blocking}
\label{sec:cache}
   Given that we have eliminated most floating-point multiplies and simulate large groups
   of diagonal gates with fast CPU instructions, performance bottlenecks shift towards memory operations. Performance is impacted by cache misses, especially when applying non-diagonal one-qubit gates (diagonal gates require a small number of linear memory passes). To improve memory locality, we take special care when forming pairs of one-qubit gates. First, the gates of each kind in a layer are sorted by the qubits they act upon. Then, we form pairs in order, so that paired up gates act on as close qubits as possible. This reduces memory strides when simulating gate pairs acting on less significant bits.
   
   A more sophisticated optimization is {\em cache blocking}.
   Rather than apply pairs of one-qubit gates in separate passes, such pairs acting on different qubits can be reordered and even applied partially in different orders. Simulating gates that act on more significant bits still suffers from long memory strides
   and frequent cache misses. To also address those gates, we developed a recursive FFT-like
   algorithm for simulating layers of one-qubit (non-diagonal) gates of a kind. Shown in 
   in Figure \ref{fig:RecursiveTransform},
   this algorithm starts with the most significant qubits and simulates gates acting on those qubits (if they exist), after which it partitions the state vector into equal-sized chunks and recurses to individual chunks.
   Chunks are chosen to have the smallest size such that the most significant one-qubit gate left can be applied within a chunk. Upon recursion, the algorithm applies multiple non-overlapping
   pairs of one-qubit gates and an occasional unpaired gate to each chunk, then moves to the next chunk. When each chunk fits in L2 cache, cache misses are reduced and performance is improved.

   \begin{figure}[htb]
%\vspace{-5mm}
\begin{lstlisting}[escapechar=@]
    
const int kRT = 8 * sizeof(@\blue{idx\_size}@) + 1;
inline int GetNextUsedQubitIndex (const @\blue{idx\_size}@ bitmask) {
    return bitmask ? __builtin_ctzl(bitmask) : kRT;
}
/* XX, XY, YX gates modify global phase. We accumulate these phases in an i-counter to avoid multiplies.*/
@\blue{idx\_size}@ XYFastTransform (
            @\blue{cmplx}@* __restrict w, //wave function
            @\blue{idx\_size}@ X_bitmask, @\blue{idx\_size}@ Y_bitmask,
            int num_qubits, int num_threads)
{
    const int idx_increment = 4; @\blue{idx\_size}@ i_phase = 0;
    if (X_bitmask & 1 || Y_bitmask & 1)
    {
        @\blue{idx\_size}@ gates_bitmask = 0;
        /* Move the bitmask so the next two qubits are the least significant and find whether to apply XX, XY, YX, YY.*/
        int gate_type = UpdateXYBitmask(
            X_bitmask, Y_bitmask, gates_bitmask,  i_phase);
        Apply2QGates(w, gate_type, gates_bitmask, num_qubits, idx_increment);
    }
    
    const int k = min(GetNextUsedQubitIndex(Y_bitmask),
                GetNextUsedQubitIndex(X_bitmask));
    // Base condition : end-case when qubit index == 65.
    // RT only supported for up to 64 bits indices here.
    if (k != kRT) {
        const @\blue{idx\_size}@ iters = 1ull << k,
        stride = 1ull << (num_qubits - k);
        X_bitmask >>= k; Y_bitmask >>= k;
        @\blue{idx\_size}@ temp_i = 0;
        for (@\blue{idx\_size}@ i = 0; i < iters ; ++i)
            temp_i += XYFastTransform( w + (i * stride), X_bitmask, Y_bitmask, num_qubits - k, num_threads);
        i_phase += temp_i / iters;
    }
    return i_phase % 4;
}
    \end{lstlisting}
    \vspace{-5mm}
    \caption{\label{fig:RecursiveTransform} Our recursive transform (RT) algorithm illustrated by applying combinations of \Xsqr and \Ysqr gates.}
\vspace{-3mm}
\end{figure}

\begin{figure}[tb]
%       \vspace{-4mm}
%      \hspace{-4mm}
       \includegraphics[width=9.4cm]{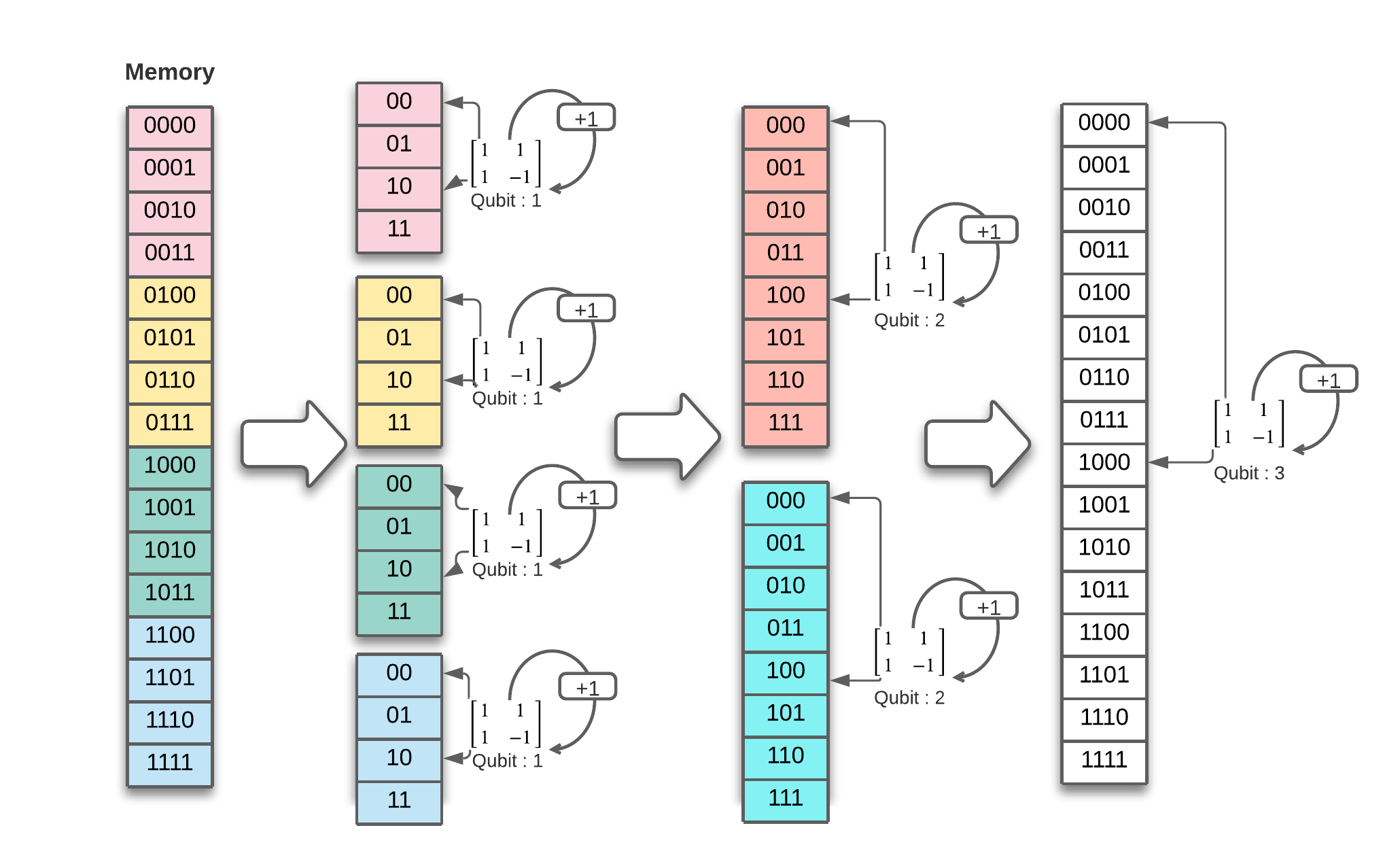}
           \vspace{-6mm}
       \caption{ \label{fig:aligned}
       When $H$ gates are simulated on qubits 1, 2 an 3, different amounts of parallelism are available to threads due to different memory strides. 
       }
%    \vspace{-2mm}
   \end{figure}
   
\subsection{Thread-level parallelism}
\label{sec:threads}
   When simulating diagonal gates, memory traversals expose significant data-level parallelism exploited by multiple CPU threads. For non-diagonal gates, the recursive FFT-like algorithm also exposes parallelism after simulating gates on the most significant qubits --- branches of recursion can be processed by different CPU threads (Figure \ref{fig:aligned}).
   This brings a 3-4$\times$ speedup with 8 threads, but for circuits with $>15$ qubits the most significant one-qubit gates become a bottleneck due to large memory strides. To simulate those gates in larger circuits,  we use a direct algorithm that partitions the state vector into regions processed by parallel threads.
   To reduce CPU cache misses, we use cache blocking and aligned-memory reads/writes described in Section \ref{sec:AVX}.
   The boundary between the more- and the less-significant qubits (used to choose between the FFT-like algorithm and direct gate application) has only a small impact on the overall runtime, so we keep these two groups as equal in size as possible.
   In C++ code, we invoke parallel threads using OpenMP pragmas, which can be disabled for sequential execution.

   % and no additional performance gains are available. 
   
\subsection{Data-level parallelism}
\label{sec:AVX}
  We achieve major performance gains with {\em aligned memory reads and writes} that fetch 256 bits of data (four complex numbers). These optimizations forced us to replace C++ STL vector classes with C-style arrays whose memory positioning can be controlled precisely. Fortunately, all large arrays in quantum circuit simulation are sized at large powers of two, and can therefore be perfectly aligned using a small amount of address arithmetic. When simulating diagonal gates, each pass becomes faster with fewer reads and writes (here we use Gray codes only within each 256-bit block). When simulating pairs of one-qubit gates, our memory traversal pattern is enhanced by cache blocking via the recursive FFT-like algorithm in Figure \ref{fig:RecursiveTransform}. Cache blocking, shown in Figure \ref{fig:aligned}, also improves temporal locality by allowing us to apply all gates that act on qubits lower than $\log_2(block\_size)$ when the block is first read for diagonal gates application - i.e., several traversals of the block are made in a short time-span. Recall that pairs of one-qubit gates are applied to four amplitudes at a time, but those four amplitudes are not contiguous in general. Therefore, we load an entire L1 cache line for each, so that four aligned reads provide data for applying two gates to four sets of amplitudes. Moreover, Section \ref{sec:reorder} clusters one-qubit gates that act
  on the same qubit(s), and we can now apply those gates
  on the same L1 cache line loaded into an AVX-2 register
  to reduce unnecessary memory transfers. Compared to 
  prior work, we do not multiply out clusters of gates as
  matrices but instead store them in a factored form.

  With entire cache lines loaded to AVX-2 registers, we made a concerted effort to leverage wide arithmetic operations from the AVX-2 instruction set. Relevant
  CPU instructions accessed through compiler intrinsics
  are listed in Table \ref{tab:AVX}. To prepare registers for AVX-2 arithmetics, 32-bit values may need to be
  shuffled using several types of bit permutations.  CPU cycles can be reduced by optimizing data shuffles and by reducing the number of arithmetic operations. The latter is facilitated by common sub-expression elimination and matrix factorizations. 
  
  \begin{figure}[tb]
    %   \vspace{-2mm}
       \centering
       \hspace{-4mm}
       \includegraphics[width=9cm]{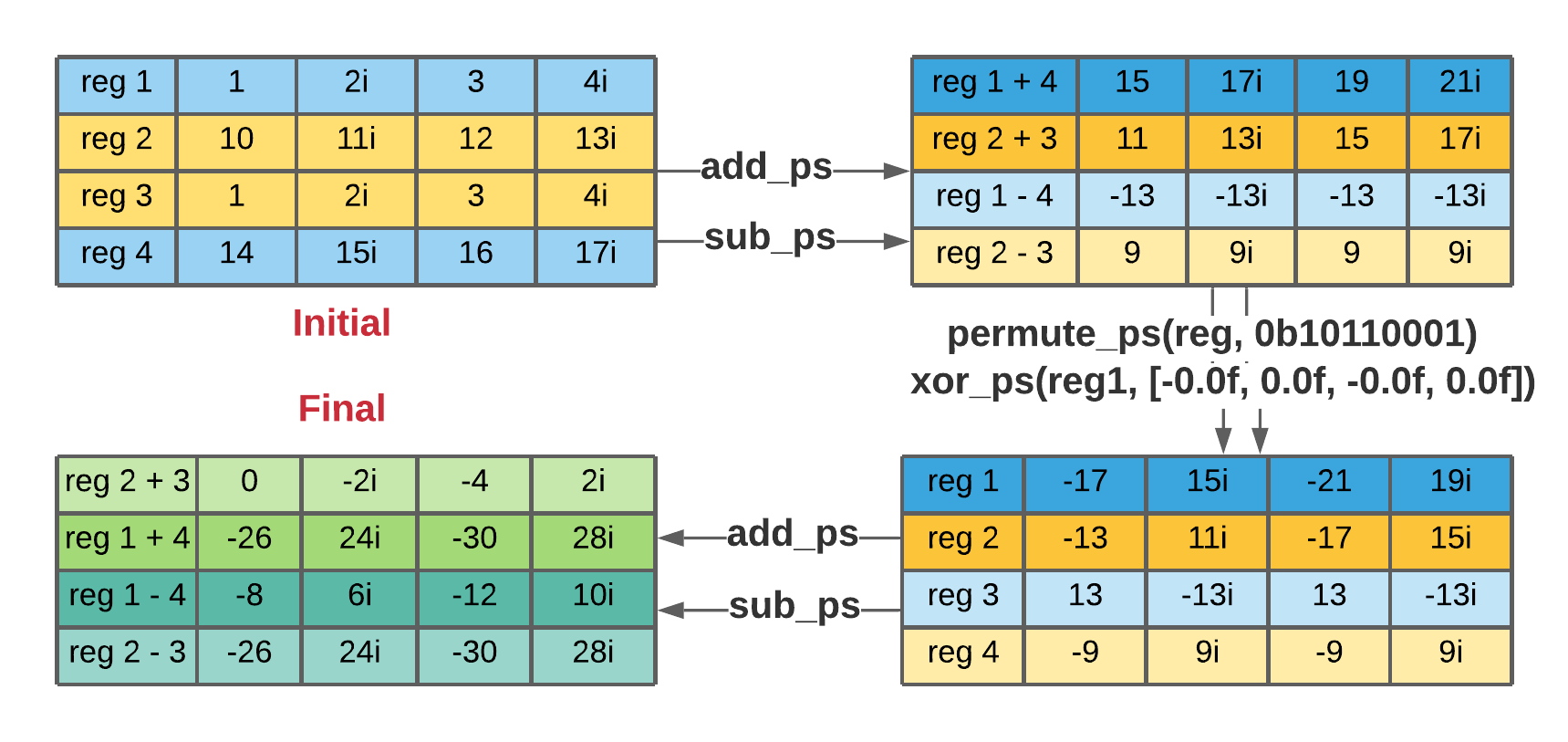}
       \vspace{-2mm}
       \caption{ \label{fig:XX}
       Applying the doubled $X\otimes X$ gate to two sets of four amplitudes stored in 256-bit registers, using
       SIMD instructions: additions, subtractions and bit swizzles. To implement multiplication by $\mathrm{i}$,
       the permute instruction switches real and imaginary components,
       then the XOR instruction negates the sign of the real component. Compiler intrinsics are explained in Table \ref{tab:AVX}.
       }
   %  \vspace{-4mm}
   \end{figure}

   \begin{example} 
    \label{ex:AVX}
    Our AVX-2 kernel to simulate
    the \Xsqr $\otimes$ \Xsqr gate (Section \ref{sec:nondiag})
    is illustrated in Figure \ref{fig:XX} and \ref{fig:XXAVX}.
    It operates on four AVX-2 registers, each holding two amplitudes
    (data supplied by aligned memory reads). 
    Multiplications by $\pm i$
    \hush{
    Figures \ref{fig:XX} and \ref{fig:XXAVX} show how to implement such multiplies} are performed in parallel by swapping the real and imaginary parts of two amplitudes and then negating their real parts
    via bitwise XOR with the {\tt -0.0f} constant. The instructions \verb| _mm256_permute_ps| and \verb|_mm256_xor_ps|
    execute in a single cycle on Intel CPUs, whereas multiplication takes 3-5 cycles depending on the CPU.
    \hush{The binary argument for \verb|_mm256_permute_ps| specifies 
    the permutation.}
    %Alternatively, amplitudes would have to be multiplied by $\pm i$ and then additional loads would be required to complete the add and subtract operations with the original amplitudes. 
    Our approach saves at least twelve CPU cycles for a pair of \Xsqr gates. 
    The add and subtract AVX-2 instructions in Figure \ref{fig:XX} \hush{and \ref{fig:XXAVX}} complete in 24 cycles. Simulating unpaired
    single \Xsqr gates requires multiplying by $1\pm i$ which we reduce to additions, subtractions, permutations and XOR operations.
    \end{example} 
 
    Customizing such permutations, arithmetic and other instructions
  to each gate type is a laborious process with careful testing. 
  To reduce CPU cycles, we investigated assembly code generated for compiler intrinsics, and these efforts were rewarded by performance benefits.
  Here we emphasize the significance of aligned memory reads/writes,
  which enable AVX-2 arithmetics and also improve bandwidth,
  keeping more CPU threads supplied with data.

\section{Simulation Comparisons}
\label{sec:empirical}

 This work targets circuits that run on NISQ computers \cite{Preskill} and are therefore limited in the number of qubits. Among such circuits, many well-known examples are fairly easy to simulate by specialized methods \cite{Viamontes}. Therefore, we focus on recent quantum-supremacy circuits from Google \cite{new_benchmarks} that
 were designed to ensure difficulty of simulation while using gates that support error correction \cite{google_2018}. The average-case difficulty of their simulation is proven analytically \cite{bouland_quantum_2018},
 and the benchmarks have been revised \cite[arxiv:1807.10749]{rr2020} to remove unintended simulation shortcuts.  Table \ref{tab:compare} shows characteristics of the benchmarks, including their large T-gate counts, which defeat stabilizer-based simulation techniques.
 
 Our methods are not limited to Google benchmarks, and our Schr\"odinger simulator
 does not exploit some of their well-known features that simplify simulation, such as their planar-grid qubit layout with nearest-neighbor qubit couplings. Therefore, one can expect comparable performance for, e.g., VQE circuits from quantum chemistry~\cite{QS_opportunities, quantumChemistry_2020}.
 In addition to pure Schr\"odinger simulation, our techniques can be used to accelerate layered simulation algorithms
\cite{rr2020,teleport2020,IBM2019storage}
 that handle a greater variety of circuits. For example, divide-and-conquer algorithms leverage Schr\"odinger simulation of $n=32$-qubit blocks to simulate $2n=64$-qubit circuits \cite{rr2020}.  Tables \ref{tab:compare} and \ref{tab:google} show results for pure Schr\"odinger and Schr\"odinger-Feynman simulation \cite{rr2020} with our optimizations included.
 
  \begin{table}[!b]
\vspace{-5mm}
\begin{center}
    \begin{tabular}{|l|c|c|} \hline
    \sc Gate simulation & \sc RR with opt & \sc RR w/o opt \\
     \sc passes & \sc \%  runtime & \sc \% runtime \\ \hline
    Initial H (32)    &  7.05  &  0.49 \\ \hline
	Unmatched final H (12) &    5.46       &  2.23 \\ \hline
	CZ \& T (300), Low  & 19.1  &  \bf 75.6 \\
	X \& Y (94) \& H (14) & & \\ \hline
	Single X (5) \& Y (1)   &  8.46  &  1.54 \\ \hline
	High X \& Y (96) \& H (6) & \bf 57.3 &  19.7 \\\hline
	Rescaling passes (2)     & 2.58  &  0.49 \\\hline \hline
	Total (560)				& \bf 72s	& \bf 2050s  \\ \hline 
    \end{tabular}
     \vspace{3mm}
    \caption{\label{tab:profile} Rollright (RR) runtime profiling data for a 32-qubit circuit, with 
    and without our optimizations. For each gate category,
    we show the number of gates simulated.}
\end{center}
\vspace{-4mm}
\end{table}

 \begin{table*}[tb]
\begin{center}
\begin{tabular}{|c||c|c|c||c|c||c|c||c|c|c||c|c||c|c|}
\hline
Circuit & \multicolumn{3}{c||}{Gates} & \multicolumn{2}{c||}{Microsoft QDK}
& \multicolumn{2}{c||}{QISkit-Aer QASM}
& \multicolumn{3}{c||}{Rollright} & \multicolumn{2}{c||}{Ratios QDK/RR}  & \multicolumn{2}{c|}{Ratios QISkit/RR}\\
depth & all & 2-q & T & time & mem & time & mem & mode & time & mem & time & mem  & time & mem \\
\small 1+26+1 & \multicolumn{3}{c||}{} & s & MiB & s & MiB & & s & MiB & \multicolumn{2}{c||}{}  & \multicolumn{2}{c|}{} \\
\hline
\multicolumn{14}{c}{{\bf MacBook Pro 2017} --- MacOS High Sierra: 16 GiB, Intel Core i7-7700HQ (2.80GHz) 4 cores 8 threads} \\
\hline \small
\verb/16q/ & 274 & 78 & 68 & 2.34 & --- & $<0.1$ & --- & S & $<0.1$ & --- & --- & --- & --- & --- \\
\hline \small
\verb/24q/ & 417 & 123 & 99 & 16.64 & 463 & 3.76 & 128.45 & S & 0.88 & 128 & \cellcolor{gray}18.91 & \cellcolor{gray}3.63 & \cellcolor{gray}4.27 & \cellcolor{gray}1.00\\
\hline \small
\verb/25q/ & 435 & 130 & 105 & 28.72 & 972 & 8.02 & 256.32 & S & 1.45 & 256 & \cellcolor{gray}19.81 & \cellcolor{gray}3.80 & \cellcolor{gray}5.53 & \cellcolor{gray}1.00 \\
\hline \small
\verb/30q/ & 524 & 161 & 119 & --- & OOM & 346.13 & 8023.39 & S & 58.8 & 8192 & --- & --- & \cellcolor{gray}5.88 & \cellcolor{gray}1.00 \\
\hline
\multicolumn{14}{c}{{ \bf Server} --- Ubuntu Linux: 144 GiB, Intel Xeon Platinum 8124M (3.00GHz) 18 cores, 72 threads} \\
\hline \small
\verb/30q/ & 524 & 161 & 119 & 1213.16 & 23959 & 18.13 & 8023.39 & \bf S & 17.6 & 8192 & \cellcolor{gray}52.29 & \cellcolor{gray}2.92 & \cellcolor{gray}1.03 & \cellcolor{gray}1.00 \\
\hline
\hline \small
\verb/30q/ & 524 & 161 & 119 & 1213.16 & 23959 & 18.13 & 8023.39 & \bf S-F & 4.23 & 27 & \cellcolor{gray}4030.43 & \cellcolor{gray}887.38 & \cellcolor{gray}4.28 & \cellcolor{gray}297.16 \\
\hline
\hline \small
\verb/32q/ & 560 & 168 & 168 & --- & OOM & 75.68 & 32022.62 & \bf S & 72.0 & 32000 & --- & --- & \cellcolor{gray}1.05 & \cellcolor{gray}1.00\\
\hline
\hline \small
\verb/32q/ & 560 & 168 & 168 & --- & OOM & 75.68 & 32022.62 & \bf S-F & 0.492 & 48 & --- & --- & \cellcolor{gray}153.82 & \cellcolor{gray}667.13\\
\hline
\hline \small
\verb/36q/ & 633 & 195 & 144 & --- & OOM & --- & OOM & \bf S-F & 90.03 & 192 & --- & --- & --- & --- \\
\hline
\end{tabular}
\parbox{18cm}{
\caption{\label{tab:compare} Comparisons of our simulator Rollright to the simulator from Microsoft QDK v0.11.2006.403 and IBM QISkit Aer v0.6.1 on benchmarks from Google (v2) \cite{GoogleSuprem,new_benchmarks} with up to 36 qubits, performed on a laptop and a midrange server. \hush{Memory usage is the increase in max resident memory versus the baseline 16-qubit simulation, to exclude code-segment size.}
\hush{In 30-qubit simulation, Rollright attains greater improvements (over Microsoft QDK) in the Schr\"odinger-Feynman mode with 18 parallel processes than in the single-process Schr\"odinger mode.
}
}}
\end{center}
\vspace{-2mm}
\end{table*}

%\vspace{-2.2mm}
 \subsection{Validation and runtime profile data}
 We implemented our algorithms in C++17 with OpenMP in a package called Rollright, compiled with Clang v11.0.3. As seen in Section \ref{sec:AVX} we use AVX-2 instructions available on commodity and server CPUs. 
 To validate simulation results
 up to 25 qubits, we saved all amplitudes of final states and checked them
 using several industry and academic simulators. For circuits with 30-36 qubits, we checked a few amplitudes with the authors of Google benchmarks.
 
 Our experiments started on a MacBook Pro 2017 laptop with 16 GiB RAM, where our baseline Schr\"odinger simulation completes a 30-qubit circuit with depth 
 $1+26+1$ in 72 s using a little over 8 GiB of RAM (1+ and +1 denote initial and final layers of Hadamard gates as in Figure \ref{fig:reorder}). Runtimes are consistent among different circuits of similar size. \hush{which makes concerns about benchmark post-selection moot.}
 On the laptop, we use a single CPU process with eight threads. For simulations on a midrange server with ample memory (Tables \ref{tab:compare} and \ref{tab:google}) we use multiple CPU processes (with eight threads each) to leverage available hardware threads.
 
Based on profiling data, I/O and circuit pre-processing take negligible time. The bottlenecks are in simulating clusters of diagonal and non-diagonal gates. We further distinguish non-diagonal gates acting on the more and the less significant qubits. To illustrate, for a $5\times6$-qubit Google circuit of depth $1+26+1$, simulating $X^{\sfrac{1}{2}}$ and $Y^{\sfrac{1}{2}}$ gates on the more significant qubits took $>75\%$ of runtime. But $\mathit{CZ}, \mathit{T}$ gates and remaining $X^{\sfrac{1}{2}}$,  $Y^{\sfrac{1}{2}}$, $\mathit{H}$ gates took $14.4\%$ runtime. 

Table \ref{tab:profile} breaks down the runtimes of a fully optimized 
(with opt) and an unoptimized (w/o opt) versions of Rollright on a 32-qubit Google circuit. Initial H gates are a separate
category because they are simulated by initializing all
amplitudes to the same value (1) and setting the carried over power of $\sqrt{2}$ accordingly. Being diagonal, 
CZ and T gates are simulated in one pass by the optimized
version. Low-qubit X, Y and H gates are simulated during
the same pass, so as to reduce memory transfers.
On each L1 cache line loaded into
an AVX-2 register by aligned memory read (and later stored
by aligmed memory write), we simulate all applicable gates. 

We see that for the unoptimized version, runtime is dominated by this joint pass (75.6\%) and the next category includes high-qubit X, Y and H gates (19.7\%). The optimized version is 28.5 times faster and significantly reduces the time taken by diagonal and low-qubit gates (19.1\%), but runtime is dominated by high-qubit X, Y and H gates (57.3\%). This is not surprising since our simulation of diagonal gates uses asymptotically fewer memory passes and exposes abundant thread-parallelism, whereas high-qubit non-diagonal gates entail multiple memory passes with large strides and limited thread parallelism. Note that we have similar numbers of low- and high-qubit X,Y gates, but high-qubit gates take much longer to simulate. Among the remaining gate categories are unpaired (single) X and Y gates, as well as a handful of unmatched trailing H gates, which take single-percent fractions of total runtime.

The apparent dominance of memory accesses over computation during optimized simulation suggests limits to further optimization. Memory throughput is also likely to become
an issue when porting our algorithms to GPUs or FPGAs.

\begin{figure*}[ht]
%\vspace{-4mm}
\begin{center}
   \hspace{-3mm}
    \begin{tabular}{ccc}
        \includegraphics[height=3.7cm,
        width=5.5cm]{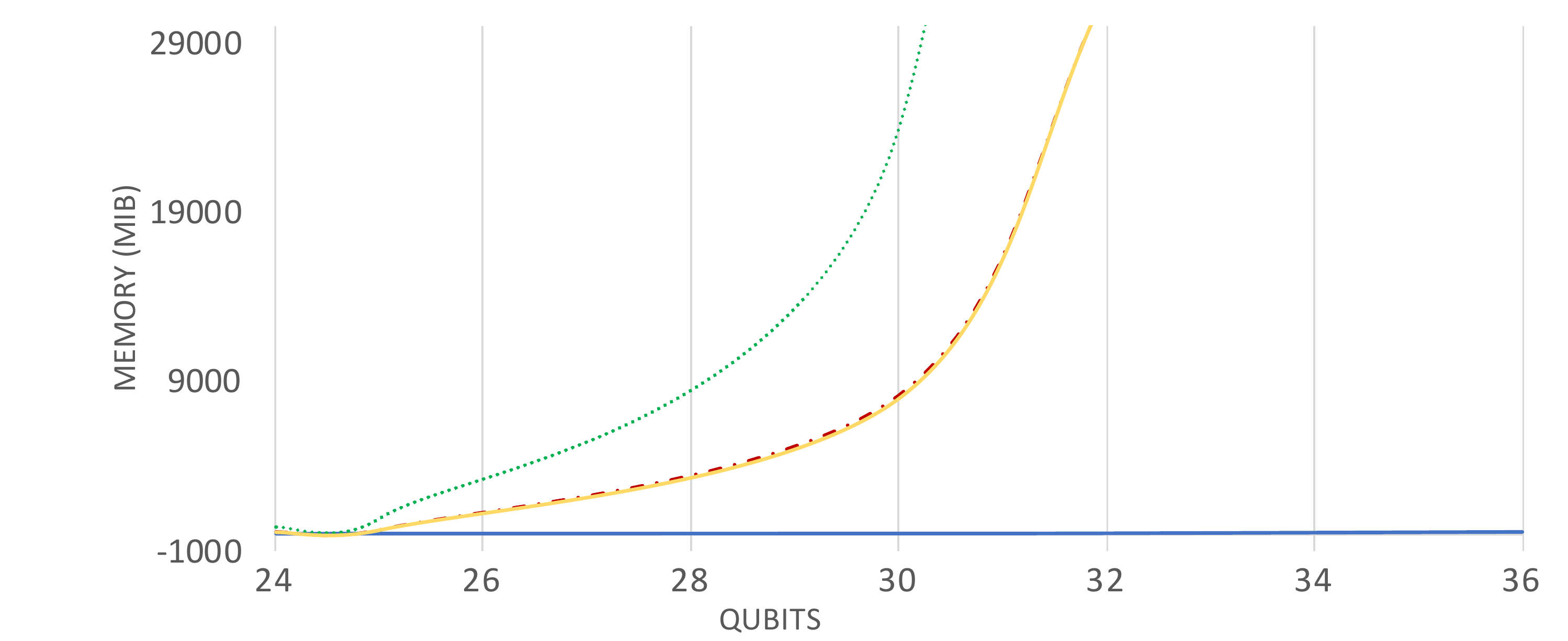} \  &
        \includegraphics[height=3.7cm,
        width=5.5cm]{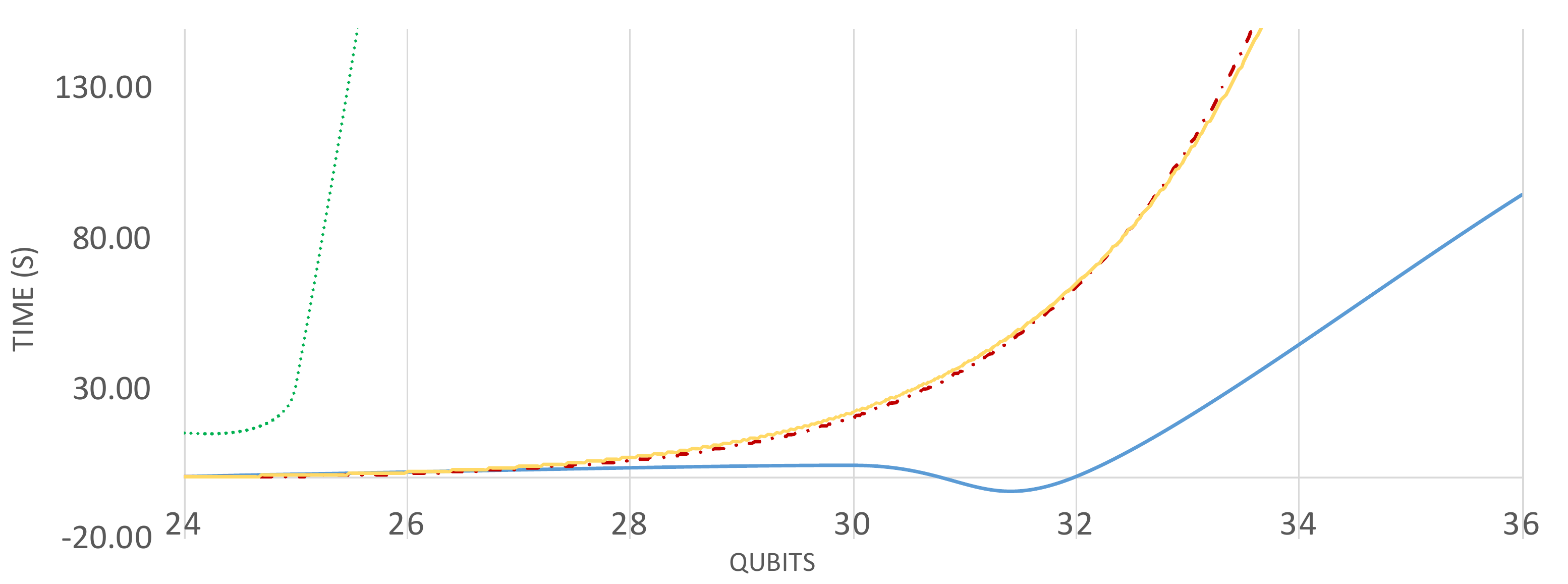} \ &
        \includegraphics[height=3.7cm,
        width=5.5cm]{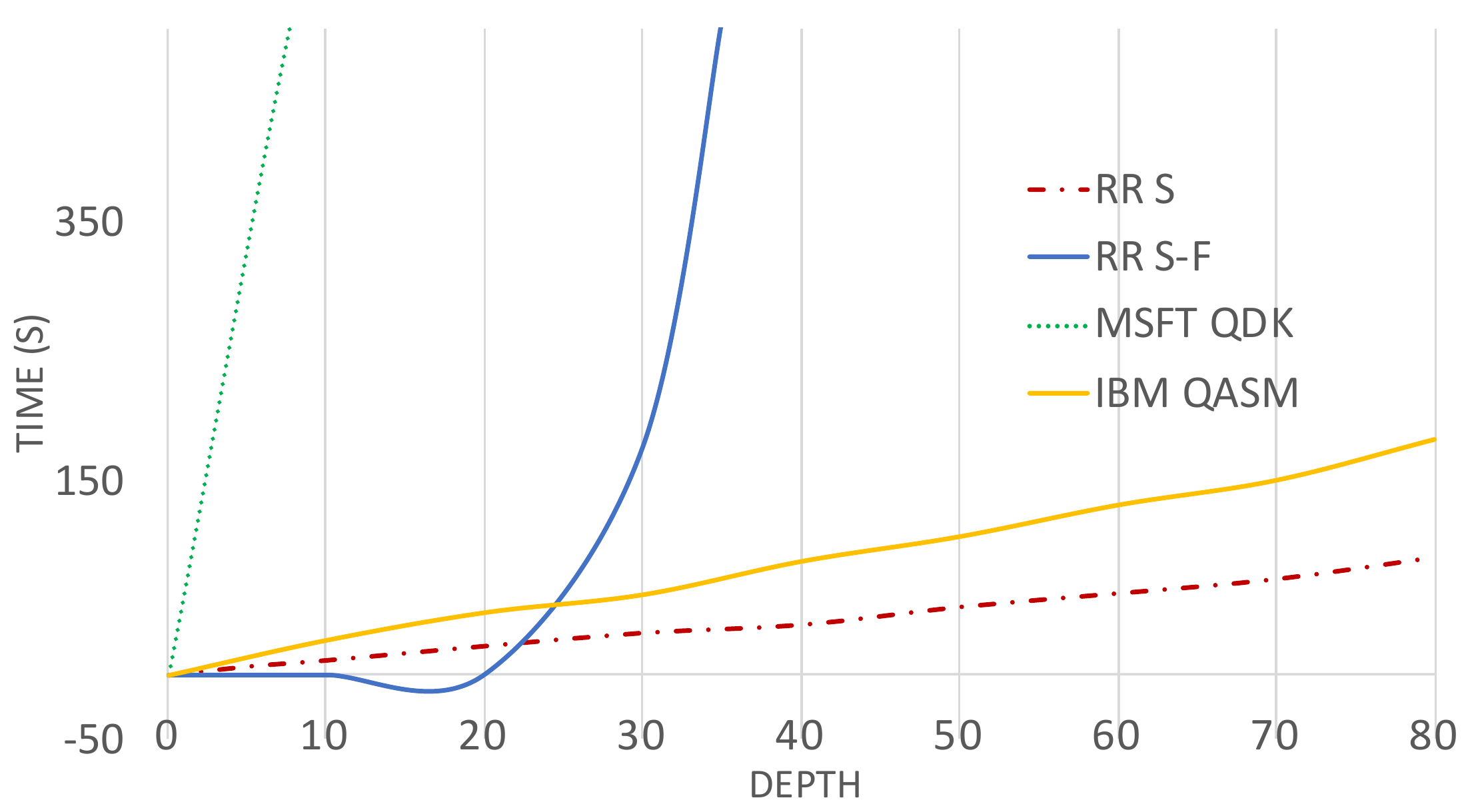} \\
    \vspace{-5mm}
    \end{tabular}
\parbox{17cm}{
\vspace{-1mm}
\caption{\label{fig:scale} Scalability simulations:
Microsoft QDK (solid green line), IBM QISkit Aer (black dashed line), as well as our simulator Rollright in Schr\"odinger (red dot-dashed line) and Schr\"odinger-Feynman (solid blue line) modes. We plot runtime and memory usage against qubit count and circuit depth. Circuit depth was varied for 30-qubit circuits. 
}
}
\end{center}
%\vspace{-2mm}
\end{figure*}

\vspace{-2mm}
\subsection{Ablation experiments}
\label{sec:ablation}
\begin{figure}[!t]
    \centering
    \includegraphics[width=9cm]{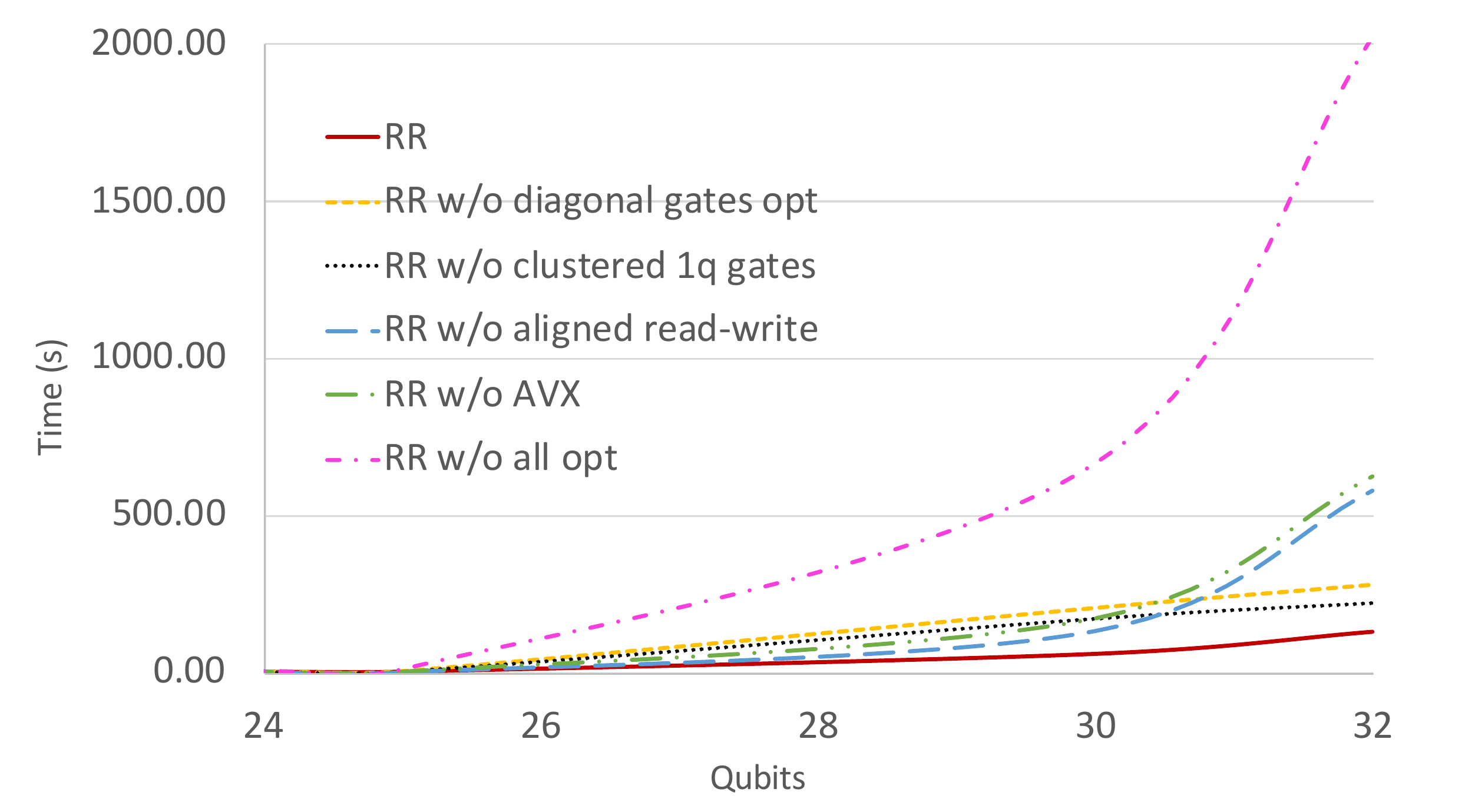}
    \vspace{-6mm}
    \caption{
    \label{fig:ablation}
    Runtime of Rollright (RR) with key features turned off one by one. The line "RR w/o all opt" shows performance without optimizations.
    }
\end{figure}

  In order to demonstrate the impact of individual optimizations on Rollright's performance, we turn them off one by one and plot the resulting runtimes in Figure \ref{fig:ablation}. In these experiments, the full configuration of Rollright uses eight threads.
  Separate lines are shown for runtime without optimizations for diagonal gates, without one-qubit gates clustered by L1 cache lines and the FFT-like algorithm,
  without AVX-2 instructions, and without aligned read-write operations. Note that without aligned read-writes, clustering by L1 cache lines makes no sense,
  and data for AVX-2 instructions are not
  readily available in AVX-2 registers. We also show runtime with all optimizations turned off --- a baselime
  Schr\"odinger simulation that applies gates one by one.
  
  Figure \ref{fig:ablation} suggests that aligned read-write and AVX instructions impact runtime by more than other individual optimizations,
  especially for circuits with over 30 qubits
  (in addition to the more efficient memory transfer,
  aligned read-write ops appear to improve CPU cache utilization). However, the cumulative impact of all
  optimizations far exceeds the impact of any one 
  optimization.

\subsection{Comparisons to Microsoft, IBM and Google}
We compared our simulator to the Microsoft Quantum Development Kit (QDK) v0.2.1806.3001 and IBM QISKit Aer v0.6.1. \hush{Microsoft QDK includes a back-end circuit simulator and front-end support for the Q\# language, integrated with Microsoft Visual Studio and available on Windows, MacOS and Linux.} Among simulations in IBM QISkit, we found QASM to be the fastest on quantum-supremacy circuits \cite{GoogleSuprem,new_benchmarks}. Table \ref{tab:compare} reports comparisons on circuits of depth $1+26+1$ with up to 36 qubits. To exclude code segments from memory comparisons, we first measured max resident memory for each simulator on the 16-qubit benchmark and then used those measurements as baselines. Memory differences among Schr\"odinger simulations, when present, are mostly due to our use of single-precision floats. Rollright's advantage in runtime is greater
and grows with the number of qubits.
\hush{and explained by two factors: ($i$) a certain slowdown due to the use of the Microsoft .NET framework, ($ii$) algorithmic differences. Only algorithmic differences explain why our runtime advantage} 

For 30-qubit circuits, the Microsoft simulator required $>$ 16 GiB memory (Rollright used a little over 8 GiB), so we also used a multicore Linux server with sufficient memory and observed that the Microsoft and IBM simulators used all available threads.
The optimizations proposed in this work apply to both Schr\"odinger and Schr\"odinger-Feynman simulation, therefore we evaluated 
Rollright in both modes. Clearly, the Schr\"odinger-Feynman simulation offers a much greater advantage in both runtime and memory on circuits of depth 1+26+1. The 32-qubit circuit uses the oblong $4\times 8$ qubit array, and the Schr\"odinger-Feynman mode of our simulator is able to exploit this shape. Therefore, we also show results for Schr\"odinger-Feynman simulation on an even-sided $6\times6$ qubit array. Our comparisons to software from IBM and Microsoft have been presented in person at these companies and helped IBM find a bug, bringing QISKit Aer memory usage down to match ours.

\begin{table}[b]
\vspace{-5mm}
\begin{tabular}{|r|c|c||c|c||c|c||c|}
\hline
  &
\multicolumn{2}{c||}{$5\times 5$ q} &
\multicolumn{2}{c||}{$6\times 5$ q} &
\multicolumn{2}{c||}{$8\times 4$ q} & 
$6\times 6$ q \\
  &
S & S-F & S & S-F & S & S-F & S-F\\
\hline
Qsim  &
1.1 & 1.64 & 24.25 & 0.47 & 152 & 0.75 & 194.12\\
\hline
RR  &
0.77 & 1.26 & 17.60 & 4.23 & 72.40 & 0.49 & 94.48 \\
\hline
ratio  &
\cellcolor{gray} 1.43 &
\cellcolor{gray} 1.30 &
\cellcolor{gray} 1.38 &
\cellcolor{gray} 0.11 &
\cellcolor{gray} 2.09 &
\cellcolor{gray} 1.53 &
\cellcolor{gray} 2.05 \\
 \hline
\end{tabular}
\caption{\label{tab:google} Server runtimes (s)
of the Google QSim simulator on Google v2
benchmarks \cite{GoogleSuprem,new_benchmarks} used in Table \ref{tab:compare}, compared to runtimes of our simulator Rollright (RR). RR runtimes for 30-36 qubits match those in the
lower half of Table \ref{tab:compare}.
}
\vspace{-5mm}
\end{table}

\begin{figure*}[!t]
\vspace{-2mm}
\begin{center}
   %\hspace{-3mm}
    \begin{tabular}{cc}
        \includegraphics[width=6.7cm]{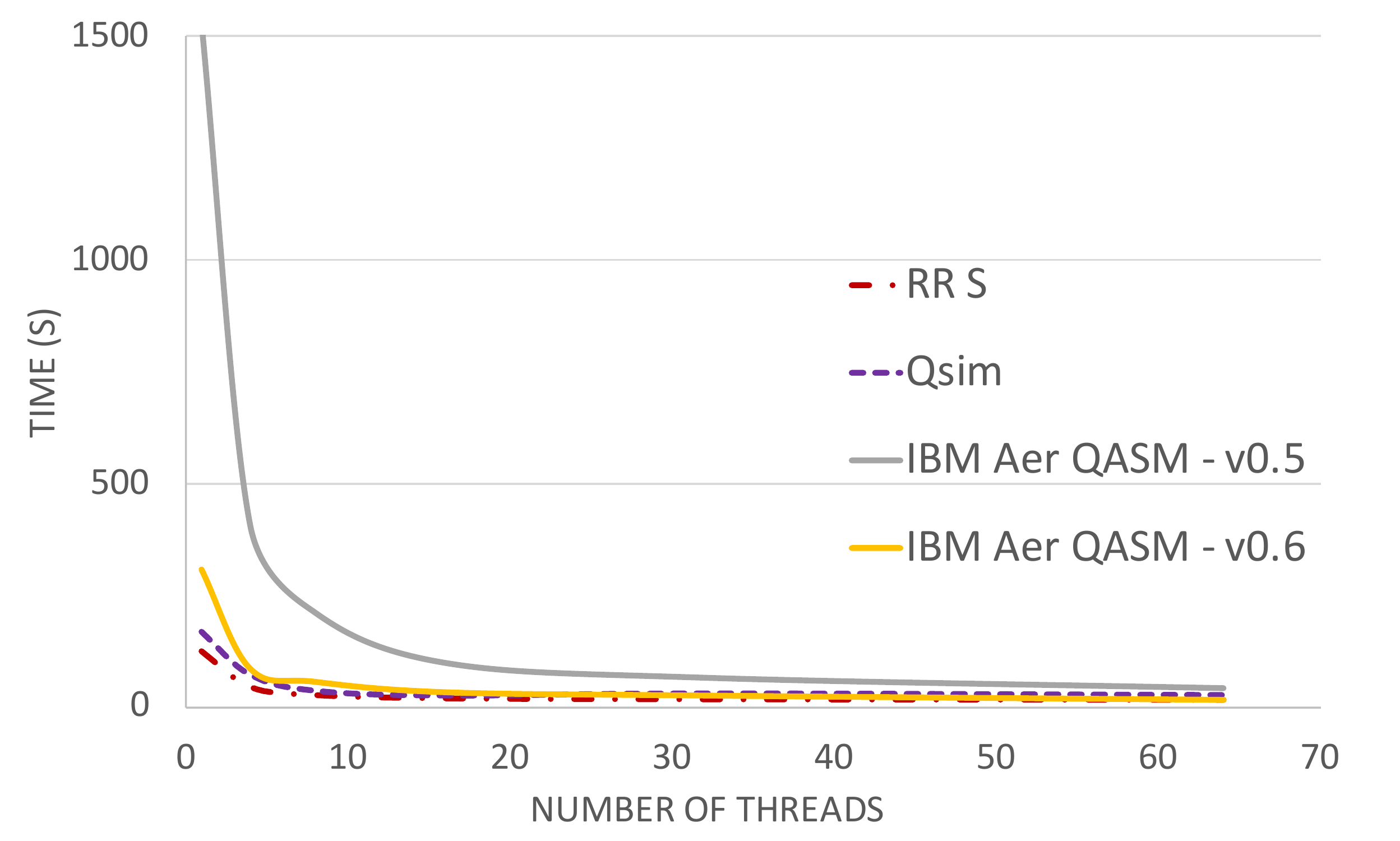} \hspace{1cm} &
        \includegraphics[width=6.7cm]{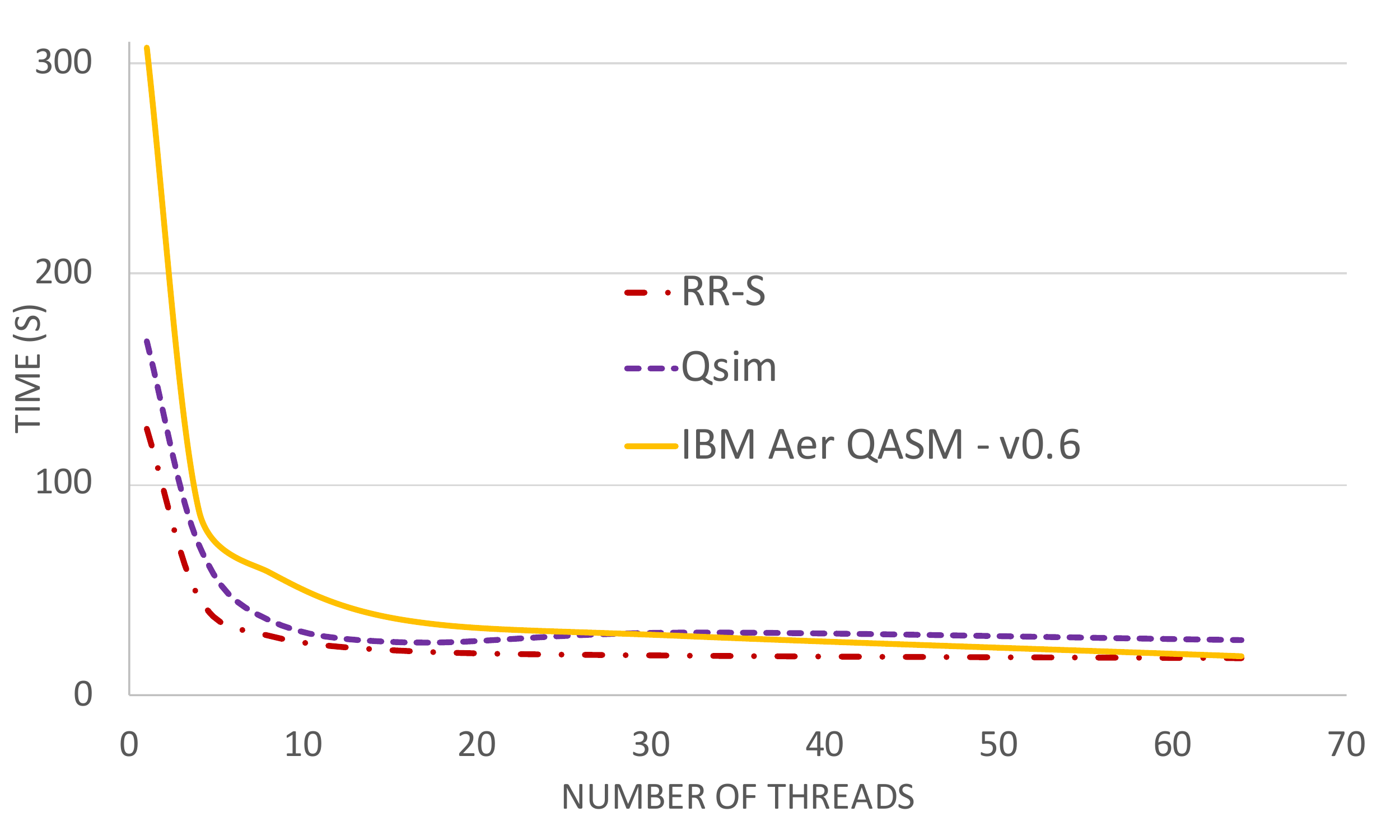} \\
    \vspace{-5mm}
    \end{tabular}
\parbox{17cm}{
\vspace{-3mm}
\caption{\label{fig:threadScale} Thread scalability: the plots of time against thread count show how IBM QISkit Aer, Google QSim and Rollright perform when simulating a 30-qubit, depth-27 circuit. The first plot highlights the dramatic improvement in performance of IBM QISkit Aer from v0.5 to v0.6. The second plot zooms in on the performance of faster simulators.}
}
\end{center}
\vspace{-2mm}
\end{figure*}

We also compared our simulator to the Qsim simulator under development at Google
({\tt https://github.com/quantumlib/qsim}).
According to the authors,
Qsim clusters one-qubit gates to nearby two-qubit gates and uses AVX-2 instructions to simulate resulting generic two-qubit gates one by one. Qsim lacks our optimizations for diagonal and one-qubit gates, as well as the FFT-like algorithm that optimizes memory access. Following our prior collaboration with Google \cite{rr2020}, Qsim supports the same simulation modes as Rollright --- Schr\"odinger and Schr\"odinger-Feynman, ---
which facilitates more detailed apples-to-apples comparisons. Runtimes in
Table \ref{tab:google} (shared with Qsim authors) were collected on the same server as for the lower half of Table \ref{tab:compare}. While Google Qsim outperforms IBM QISkit and Microsoft QDK on Google benchmarks, Rollright remains ahead, confirming the impact of our proposed methods. 

% \vspace{-3mm}
\subsection{Scalability studies and use models}
\label{sec:scalability}

Table \ref{fig:threadScale} shows simulator performance for varying numbers of CPU threads. Different simulators exhibit similar scaling, with Rollright remaining ahead in all cases.
The results in Table \ref{tab:compare} show massive advantage of Schr\"odinger-Feynman simulation, but pure Scr\"odinger simulation remains attractive for deep quantum circuits, e.g., in quantum chemistry applications \cite{QS_opportunities, quantumChemistry_2020} and/or when supercomputing resources are available \cite{de_raedt_massively_2007,wecker2014liquid,de_raedt_massively_2018,PB,qHIPSTER,IBM2017breaking,Taihu,IBM2019storage}.
$2^\mathrm{num\_qubits}$ and runtime as
$\mathrm{depth}\times 2^\mathrm{num\_qubits}$. 
Simulating 40-qubit circuits this way would require servers with $>$8 TiB (now available from Microsoft Azure and Amazon AWS). 
In the Schr\"odinger-Feynman mode, 
memory usage can be kept low (see details in \cite{rr2020}) by serializing the computation, but if massive parallel resources are available, using greater peak memory can decrease the latency of simulation. In the meantime, runtime grows as $2^{\mathrm{num\_qubits}/2 + \mathrm{depth}/C}$ for a large $C>0$. 
Figure \ref{fig:scale} uses larger
supremacy benchmarks from Google to
illustrate these differences between Schr\"odinger and Schr\"odinger-Feynman simulation by plotting memory usage and runtime for varied qubit counts and circuit depth. The linear runtime of Schr\"odinger simulation vs. circuit depth (regardless of gate typess) improves upon the semi-exponential scaling of Schr\"odinger-Feynman simulation.

Distributed Schr\"odinger-Feynman simulations with Rollright in Google Cloud \cite{rr2020} show that our methods are relevant to boundeded-depth 56- and 64-qubit circuits. Our methods also fit in
unbounded-depth Schr\"odinger simulations on supercomputers~\cite{Taihu}. Results in \cite{teleport2020} cast the simulation of shallow wide circuits to that of deep narrow circuits, where pure Schr\"odinger simulation does well. \hush{We also believe that some of our algorithms --- essentially fast matrix multiplications --- generalize to tensors and can be adapted to tensor-network contraction \cite{Villalonga_2020}.
}

\section{Conclusions}
\label{sec:conclusions}
  {\em Near-term intermediate-scale quantum} (NISQ) computers \cite{Preskill} are operating with $<$64 qubits in 2020.
  \hush{including a recent demonstration of quantum-computational supremacy by Google \cite{Arute2019} with 53 qubits.}
  Quantum circuits running on such computers support many science experiments \cite{QS_opportunities} and motivate circuit optimization tasks, which often require simulation on conventional computers. Recent advances in quantum chemistry offer synthesis methods for NISQ circuits that model molecular configurations and compute their energy levels. Here quantum-circuit simulation is needed to develop and validate advanced quantum technologies, such as quantum-on-quantum simulators~\cite{QS_opportunities}.%
 
  Among the many simulation algorithms, this work focuses on Schr\"odinger simulation that can be used independently
  or in layered simulation algorithms
\cite{rr2020,teleport2020,IBM2019storage}
 that handle a greater variety of circuits.
   Our algorithmic optimizations collectively provide hefty speed-ups over quantum simulators from Microsoft and IBM on hard circuits from Google. These speedups are not limited by a constant factor, but grow with the number of qubits.
  Our high-level and some low-level optimizations --- gate clustering by type, aligned memory reads/writes, gate clustering by cache line, fast simulation of diagonal gates, the recursive FFT-like algorithm,  --- are generic. Low-level optimizations are tuned to gates that support quantum errror correction, are
  available on Google chips and are used in Google benchmarks~\cite{GoogleSuprem,new_benchmarks}. 
  Additional gates can be supported natively or by expressing them in terms of native gates.
  \hush{A particularly remarkable example is the Controlled-Phase gate supported by Google's Sycamore chip \cite{Arute2019}. Appropriate gate decompositions can be obtained using the closely related iSWAP gate which can be expressed using a $CZ$ gate,
  a SWAP gate and one-qubit gates \cite{nielsen2003quantum}. SWAP gates
  can be bubbled up through the entire circuit and absorbed by permutations of
  input qubits. Therefore, SWAP gates do not materially
  affect the overall runtime of simulation.}
  
  For evaluation, we use medium-size circuits which most industry simulators can handle today, but our contributions help with many more qubits as shown in \cite{rr2020, teleport2020} and directly benefit
  supercomputing simulations \cite{Taihu}.
  Pure Schr\"odinger simulation is well-suited for deep circuits for VQE algorithms in quantum chemistry \cite{QS_opportunities, quantumChemistry_2020}.
  \hush{
  However, supercomputers do not support the needs of frequent on-demand simulation
  during practical work with quantum algorithms,
  quantum error-correcting code, and/or NISQ hardware. For such uses, better approaches include smaller-scale Schr\"odinger simulation on single high-mem servers and distributed Schr\"odinger-Feynman simulation 
  in commercial clouds \cite{rr2020}.
  }

% \begin{figure}[h]
%  \centering
%  \includegraphics[width=\linewidth]{sample-franklin}
%  \caption{1907 Franklin Model D roadster. Photograph by Harris \&
%    Ewing, Inc. [Public domain], via Wikimedia
%    Commons. (\url{https://goo.gl/VLCRBB}).}
%  \Description{A woman and a girl in white dresses sit in an open car.}
%\end{figure}

%\noindent
%{\bf Acknowledgments.}
\begin{acks}
 We thank Dmitri Maslov, Sergio Boixo and Sergei Isakov for insightful comments.
\end{acks}

%\newpage
\bibliographystyle{ACM-Reference-Format}
\bibliography{sample-base}

\hush{
\bibitem{64q}
\bibinfo{author}{Chen, Z.-Y.} \emph{et~al.}
\newblock \bibinfo{title}{64-qubit quantum circuit simulation}.
\newblock \emph{\bibinfo{journal}{arXiv:1802.06952}}  (\bibinfo{year}{2018}).
}

%appendix

% \newpage
% \appendix
%\section{Representative codes}

\end{document}